\documentclass[twocolumn,trackchanges]{aastex631}

\newcommand{\Msun}{{\mathrm  M_{\odot}}}
\newcommand{\Zsun}{{\mathrm  Z_{\odot}}}
\newcommand{\SNII}{\mbox{CCSNe}}
\newcommand{\PopIII}{\mbox{Pop I\hspace{-1pt}I\hspace{-1pt}I}}
\newcommand{\PopII}{\mbox{Pop I\hspace{-1pt}I}}

\newcommand{\ts}{t_\mathrm{dep}}
\newcommand{\ti}{t_\mathrm{in}}
\newcommand{\fo}{f_\mathrm{out}}
\newcommand{\Msms}{M_\mathrm{SMS}}

\newcommand{\Gyr}{{\rm Gyr}}

\graphicspath{{./}{figures/}}

\begin{document}

\title{Probing Chemical Enrichment in Extremely Metal-Poor Galaxies}

\author[0000-0002-5045-6052]{Keita Fukushima}
\affiliation{Theoretical Astrophysics, Department of Earth and Space Science, Graduate School of Science, Osaka University, 1-1 Machikaneyama, Toyonaka, Osaka 560-0043, Japan}
\affiliation{Waseda Research Institute for Science and Engineering, Faculty of Science and Engineering, Waseda University, 3-4-1 Okubo, Shinjuku, Tokyo 169-8555, Japan}
\affiliation{Institute for Data Innovation in Science, Seoul National University, Seoul 08826, Republic of Korea}

\author[0000-0001-7457-8487]{Kentaro Nagamine}
\affiliation{Theoretical Astrophysics, Department of Earth and Space Science, Graduate School of Science, Osaka University, 1-1 Machikaneyama, Toyonaka, Osaka 560-0043, Japan}
\affiliation{Theoretical Joint Research Project, Forefront Research Center, Graduate School of Science, Osaka University, 1-1 Machikaneyama, Toyonaka, Osaka 560-0043, Japan}
\affiliation{Kavli IPMU (WPI), The University of Tokyo, 5-1-5 Kashiwanoha, Kashiwa, Chiba, 277-8583, Japan}
\affiliation{Department of Physics \& Astronomy, University of Nevada, Las Vegas, 4505 S. Maryland Pkwy, Las Vegas, NV 89154-4002, USA}
\affiliation{Nevada Center for Astrophysics, University of Nevada, Las Vegas, 4505 S. Maryland Pkwy, Las Vegas, NV 89154-4002, USA}

\author{Akinori Matsumoto}
\affiliation{Institute for Cosmic Ray Research, The University of Tokyo, 5-1-5 Kashiwanoha, Kashiwa, Chiba 277-8582, Japan}
\affiliation{Department of Physics, Graduate School of Science, The University of Tokyo, 7-3-1 Hongo, Bunkyo, Tokyo 113-0033, Japan}

\author[0000-0001-7730-8634]{Yuki Isobe}
\affiliation{Kavli Institute for Cosmology, University of Cambridge, Madingley Road, Cambridge, CB3 0HA, UK}
\affiliation{Cavendish Laboratory, University of Cambridge, 19 JJ Thomson Avenue, Cambridge, CB3 0HE, UK}
\affiliation{Waseda Research Institute for Science and Engineering, Faculty of Science and Engineering, Waseda University, 3-4-1, Okubo, Shinjuku, Tokyo 169-8555, Japan}

\author[0000-0002-1049-6658]{Masami Ouchi}
\affiliation{National Astronomical Observatory of Japan, National Institutes of Natural Sciences, 2-21-1 Osawa, Mitaka, Tokyo 181-8588, Japan}
\affiliation{Institute for Cosmic Ray Research, The University of Tokyo, 5-1-5 Kashiwanoha, Kashiwa, Chiba 277-8582, Japan}
\affiliation{Kavli Institute for the Physics and Mathematics of the Universe (WPI), University of Tokyo, Kashiwa, Chiba 277-8583, Japan}

\author[0000-0001-8226-4592]{Takayuki R. Saitoh}
\affiliation{Department of Planetology, Graduate School of Science, Kobe University, 1-1 Rokkodai-cho, Nada-ku, Kobe, Hyogo 657-8501, Japan}
\affiliation{Center for Planetary Science (CPS), Graduate School of Science, Kobe
University 1-1 Rokkodai, Nada-ku, Kobe, Hyogo 657-8501, Japan}

\author[0000-0002-5661-033X]{Yutaka Hirai}
\affiliation{Department of Community Service and Science, Tohoku University of Community Service and Science, 3-5-1 Iimoriyama, Sakata, Yamagata 998-8580, Japan}



\begin{abstract}

The chemical composition of galaxies offers vital insights into their formation and evolution. In particular, the relationship between helium abundance (He/H) and metallicity serves as a key diagnostic for estimating the primordial helium yield from Big Bang nucleosynthesis.
We investigate the chemical enrichment history of low-metallicity galaxies, focusing especially on extremely metal-poor galaxies (EMPGs), 
using one-zone chemical evolution models. 
Adopting elemental yields from \citet{Limongi2018ApJS}, our models reach $\mathrm{He/H} \sim 0.089$ at $(\mathrm{O/H}) \times 10^5<20$, yet they fall short of reproducing the elevated $\mathrm{He/H}$ values observed in low redshift dwarf galaxies.  
In contrast, the observed Fe/O ratios in EMPGs are successfully reproduced using both the \citet{Nomoto13} and \citet{Limongi2018ApJS} yield sets.
To address the helium discrepancy, we incorporate supermassive stars (SMSs) as {\PopIII} stars in our models.  We find that SMSs can significantly enhance $\mathrm{He/H}$, depending on the mass-loss prescription.  When only $10\%$ of the SMS mass is ejected, the model yields the steepest slope in the $(\mathrm{O/H}) \times 10^5$ –- He/H relation.  Alternatively, if the entire outer envelope up to the CO core is expelled, the model can reproduce the high He/H ratios observed in high-redshift galaxies ($\mathrm{He/H} > 0.1$). 
Additionally, these SMS-enriched models also predict elevated N/O ratios, in agreement with recent JWST observations of the early universe.

\end{abstract}

\keywords{galaxies: abundances --- galaxies: dwarf --- galaxies: high-redshift --- galaxies: evolution --- methods: numerical}


\section{Introduction} \label{sec:intro}

Galaxy evolution begins with the formation of the first galaxies, which also mark the earliest sites of metal enrichment in the universe. 
Advanced cosmological hydrodynamic simulations suggest that these primordial systems, forming at redshifts $z>10$, are characterized by extremely low metallicities ($Z=0.01-0.001\,\Zsun$) and  relatively small stellar masses ($M_\star\lesssim10^6\,\Msun$)
\citep{Wise2012, Johnson2013MNRAS, Kimm_Cen2014ApJ,Romano-Diaz2014ApJ, Yajima2017ApJ, Yajima2023MNRAS}.

The James Webb Space Telescope (JWST) has spectroscopically confirmed galaxies at $z>10$ \citep{Roberts-Borsani2023Natur, Williams2023Sci, Curtis-Lake2023NatAs, Bunker2023A&A, Arrabal2023ApJ, Arrabal2023Natur, Harikane_2023_c_arXiv, Harikane2023a_ApJS}, enabling the first investigations of the mass--metallicity relation (MZR) at such high redshifts \citep{Curti2023MNRAS.518, Nakajima2023arXiv}. 
Among these, GN-z11 --- remarkable for its substantial stellar mass despite its early cosmic age --- has attracted particular attention and became the focus of several in-depth observational studies \citep{Cameron2023MNRAS, Bunker2023A&A, Senchyna2023arXiv, Isobe2023ApJ}.

Nevertheless, direct observations of low-mass galaxies in the early universe remain challenging without the aid of gravitational lensing.
\citet{Isobe2022ApJ} showed that H$\alpha$ emission from galaxies with stellar masses around $M_\star \sim 10^6\,\Msun$ can only be detected up to $z<1$ with JWST, and up to $z<2$ with next-generation facilities such as the Thirty Meter Telescope (TMT), in the absence of lensing amplification \citep[c.f.][]{Vanzella2023A&A}.

Extremely Metal-Poor Galaxies (EMPGs) are considered promising local analogs of low-mass first galaxies and may provide critical insights into the physical properties and evolutionary pathways of primordial galaxies.
EMPGs are characterized by low stellar masses ($M_\star<10^7\,\Msun$), low metallicities ($Z<0.1\,\Zsun$), and high specific star formation rates (sSFR$\sim100\,\Gyr^{-1}$), mirroring the expected features of the earliest  galaxies \citep{Kojima2020ApJ}.
\citet{Curti2024A&A...684A..75C} examined the MZR for low-mass galaxies at $3<z<10$, observed via gravitational lensing, and found it comparable to that of nearby low-metallicity starburst systems such as the so-called  
``Blueberry" galaxies \citep{
Yang2017ApJ_Blueberry}.
Thus, EMPGs offer a unique observational window into the formation and early evolution of galaxies in the high-redshift universe. 

Despite their significance, the chemical enrichment history of EMPGs remains poorly understood. 
Observations reveal that some EMPGs exhibit elevated Fe/O approaching solar values \citep{Izotov2018MNRAS_J0811+4730, Kojima2021ApJ}. 
Some chemical evolution models \citep{Isobe2022ApJ, Watanabe2023} suggest that such enrichment may require contributions from energetic core-collapse events, such as hypernovae and/or pair-instability supernovae (PISNe), associated with massive stars \citep{Barkat1967PhRvL, Heger_Woosley_2002ApJ, UmedaNomoto2002ApJ, Nomoto13}. 
Notably, galaxies with enhanced Fe abundance ([Fe/O] $= 0.3$) have been observed as early as $z=10.60$, with models invoking PISNe and bright hypernovae to explain such features \citet{Nakane2024ApJ}.
These findings are especially relevant given the expectation that young, low-metallicity galaxies may form stars under a top-heavy initial mass function (IMF) \citep[e.g.][]{Kumari2018MNRAS, Zou2024arXiv240200113Z, Chon2021MNRAS, Chon2022MNRAS, Chon2023arXiv231213339C}.

The helium-to-hydrogen abundance ratio (He/H), when examined as a function of metallicity, offers a potential means of estimating the primordial helium abundance. However, this relationship remains highly uncertain \citep{Matsumoto2022ApJ}. 
\citet{Vincenzo2019A&A} investigated the He/H -- metallicity (12+$\log$(O/H)) relation using both one-zone models and cosmological chemodynamical simulations. 
They compared \citet{Nomoto13} and \citet{Limongi2018ApJS} yield models for core-collapse supernovae ({\SNII}), and \citet{Karakas10} and \citet{Ventura2013MNRAS} for asymptotic giant branch (AGB) stars. 
Their results demonstrated that He/H can be significantly elevated when using the \citet{Limongi2018ApJS} yields, which account for the effects of Wolf-Rayet stars at low metallicities. 

In addition to conventional sources, supermassive stars (SMSs) have been proposed as a possible origin of enhanced He/H in young galaxies.
\citet{Yanagisawa2024ApJ} reported three galaxies at $z=5.92, 6.11$, and 6.23 with exceptionally high helium abundances (He/H $>0.1$) at $(\mathrm{O/H})\times10^5 < 7$. 
These values exceed those typically observed in nearby dwarf galaxies \citep{Hsyu2020ApJ} and EMPGs \citep{Matsumoto2022ApJ}.
SMSs have also gained attention as a potential explanation for chemically peculiar objects such as GN-z11, which exhibits a high N/O ratio at early cosmic times \citep{Charbonnel2023A&A, Isobe2023ApJ, Nandal2024AA}.

In this study, we aim to elucidate the formation and chemical evolution of EMPGs and high-$z$ galaxies through a series of one-zone chemical evolution models. These models are designed to probe how fundamental parameters --- such as stellar yields, star formation history, and gas inflow/outflow --- affect the chemical evolution of galaxies with low metallicities and young stellar populations.  By exploring different nucleosynthetic yield sets, we assess the extent to which these factors influence the enrichment of helium and other elements in early galaxies.

This paper is organized as follows. 
In Section~\ref{sec:method}, we describe the methodology, focusing on the one-zone chemical evolution model. 
Section~\ref{subsec:res_onezone_standard} examines the effects of metallicity on the He/H and Fe/O ratios, using two different sets of stellar yield models. 
In Section~\ref{subsec:SMSEnrichment}, we explore the impact of SMSs and variations of star formation history on chemical enrichment within the one-zone framework.
Section~\ref{subsec:discussion_SMS_starburst} discusses the implications for nitrogen enrichment, particularly the $[\mathrm{N/O}]$ ratio,  
while Section~\ref{subsec:How_to_estimate} addresses how the primordial helium abundance can be inferred from observations of high-$z$ and low-metallicity galaxies.
Our conclusions are summarized in Section~\ref{sec:summary}.

Appendix~\ref{sec:appendix_One-zone} provides further details of the one-zone model implementation.  
Appendix~\ref{sec:appendix_CELib} presents the elemental yield data used in this study, including models from \citet{Nomoto13, Limongi2018ApJS, Nandal2024AA}.

Throughout this paper, we adopt the following solar abundance ratios: $ \log(\mathrm{Fe/O}) = -1.23 $ and $ \log(\mathrm{N/O}) = -0.86 $ \citep{Asplund2021AandA}. 
Elemental abundance ratios are expressed relative to solar values and defined as $[\mathrm{A/B}] = \log_{10} \left( (N_\mathrm{A} / N_{\mathrm{A},\odot})/(N_\mathrm{B} / N_{\mathrm{B},\odot}) \right)$, 
where $ N_\mathrm{A} $ and $ N_\mathrm{B} $ are the numbers of elements A and B, respectively, and the subscript $\odot$ denotes their solar values.
For He/H, we use the ratio $ \mathrm{He/H} = N_\mathrm{He} / N_\mathrm{H} $, and oxygen abundance is also expressed as $\mathrm{(O/H)\times10^5} = N_\mathrm{O}/N_\mathrm{H} \times 10^5$.

\section{Method} \label{sec:method}

We follow the one-zone box model framework of \citet{Kobayashi_Taylor_2023arXiv} as our standard approach;  further details are provided in Appendix~\ref{sec:appendix_One-zone}. 
In this model, the outflow rate is computed based on the energy injected by stellar feedback, with the energy release rates derived using the {\sc CELib} code \citep{Saitoh16, Saitoh17}. 
To explore a wide range of evolutionary scenarios, we perform 320 model runs, systematically varying key parameters.
These include the gas-depletion timescale $\ts$, gas inflow timescale $\ti$, gas outflow rate $\fo$, and the fraction of metals in the inflowing gas $f_\mathrm{inf}$.
A summary of the parameter space explored in our models is presented in Table~\ref{tab:parameters} for reference and clarity.

Chemical evolution is computed using the {\sc CELib} code \citep{Saitoh16, Saitoh17}, which incorporates the effects of {\SNII}, type Ia supernovae (SN Ia), and AGB stars.
We adopt the {\SNII} yield from \citet{Nomoto13}, the SN Ia yields from the N100 model of \citet{Seitenzahl2013MNRAS}, the AGB star yields from \citet{Karakas10}, and the super-AGB star yields from \citet{Doherty2014MNRAS}. The \citet{Chabrier03} IMF is adopted with a stellar mass range of 0.1--100\,$\Msun$.

We also perform an additional set of calculations using the \citet{Limongi2018ApJS} yields for CCSN, specifically the ``set R" model, in which all stars more massive than 25\,$\Msun$ are assumed to fully collapse into black holes, as an alternative to the \citet{Nomoto13} yields, facilitating a comparative analysis.
Limongi's set R yields include three models corresponding to different stellar rotation velocities, and {\sc CELib} selects among them based on an empirical relation between stellar mass and rotation rate \citep{Prantzos2018MNRAS}. 
For Limongi yield at [Fe/H] $< -3$, the {\sc CELib} code applies the yield value at [Fe/H]$ = -3$.
Since the Limongi yield table covers up to $m = 120 M_{\odot}$, we also adopt $120 M_{\odot}$ as the maximum stellar mass in the {\PopIII} top-heavy IMF \citep{Susa14}. 
Hereafter, we refer to the calculation using the \citet{Nomoto13} yield for CCSN as Model-N, and the calculation using the \citet{Limongi2018ApJS} yield as Model-L (see Table~\ref{tab:modelsets} for a summary).

We treat stars with $Z\leq10^{-5}\,Z_\odot$ as {\PopIII} and adopt the following yield prescriptions: {\SNII} yields from \citet{Nomoto13}, AGB star yields from \citet{Campbell_Lattanzio2008AA} and \citet{Gil-Pons2013AA}, a top-heavy IMF from \citet{Susa14}, spanning a stellar mass range of 0.7--300\,$\Msun$. 
The {\SNII} yields for {\PopIII} stars also include the contribution from PISN \citep{Nomoto13}. 
A delay-time distribution function with a power law of $t^{-1}$ was used for the SN Ia event rate \citep{Totani08, Maoz12, Maoz2014}, which is turned on after $4\times10^7\,\mathrm{yr}$.
The hypernovae mixing fraction $f_\mathrm{HN}$ is set to 0.05.
Further details of the adopted yield tables and implementation are provided in \citet{Fukushima23MN}.

\begin{table}
\begin{center}
\caption{Summary of our one zone model parameters. See Appendix~\ref{sec:appendix_One-zone} for the definition of the parameters. 
}
\label{tab:parameters}
\begin{tabular}{cccc}
\hline \hline
\( {\ts}\) (yrs) & \( {\ti} \) (yrs) & \( {\fo} \) & \( f_{\mathrm{inf}} \) \\
\hline
\( 10^7 \) & \( 10^7 \) & 0.0 & 0.0 \\
\( 10^8 \) & \( 10^8 \) & 0.1 & 0.01 \\
\( 10^9 \) & \( 10^9 \) & 1.0 & 0.1 \\
\( 10^{10} \) & \( 10^{10} \) & 10 & 1.0 \\
 -- & -- & 100 & -- \\
\hline
\end{tabular}
\end{center}
\end{table}

In our standard model, the star formation rate (SFR) is defined as the gas mass divided by the depletion time $\ts$ (see eq.~\ref{eq:method_SFR}). Because the governing equations are expressed in terms of gas mass fractions, the derived SFR has units of [1/yr], representing the inverse of the $\ts$.

To explore the potential impact of {\PopIII} stars being SMSs, we implement an intermittent star formation model, in which an SMS of mass $\Msms$ forms only when the total accumulated stellar mass at $Z_\star < 10^{-5}\Zsun$ exceeds $\Msms$. During this accumulation phase, conventional star formation is suppressed, and stellar mass growth proceeds solely through sporadic SMS formation. For SMS yields, we adopt the $\Msms=6127\,\Msun$ model from \citet{Nandal2024AA}.
We consider two mass-loss scenarios for the SMS: 
(1) Model-N\&N-10$\%$, where only the outermost $10\%$ of the SMS mass is ejected, and 
(2) Model-N\&N-CO, where the entire envelope outside the CO core is expelled. 

In addition, we examine versions of both models in which the SMS directly collapses (DC) into a black hole, retaining the same yields. These are denoted Model-N\&N-10\%-DC and Model-N\&N-CO-DC, respectively.

For comparison, we also perform a calculation using the same intermittent star formation framework, but adopting the \citet{Nomoto13} yield.
This model, referred to as Model-N$_\mathrm{int}$, shares the same star formation prescription as the Model-N\&N series: no stars are formed until enough gas accumulates to form a $6127\,\Msun$ star. The yield used for  this model is taken from \citet{Nomoto13} for $Z_\star < 10^{-5}\Zsun$.

A summary of the model variants --- including yield sets, star formation modes, and treatment of direct collapse --- is provided in Table~\ref{tab:modelsets}. 
When stellar metallicity is the same, we adopt the same yields for AGB stars and SNe Ia across models. 
However, in the Model-N\&N series, which all stars at $Z_\star < 10^{-5}\Zsun$ are assumed to be SMSs, we omit AGB and SNIa contributions, as these are not applicable to SMS-dominated stellar populations.

\begin{table*}
\begin{center}
\caption{
 Summary of yield models used in this paper.\\ 
{\footnotesize $\dagger$ For $Z_\star < 10^{-5}\,\Zsun$: AGB yield is from \citet{Campbell_Lattanzio2008AA, Gil-Pons2013AA}; SNIa yield is from \citet{Seitenzahl2013MNRAS}. \\
$\sharp$ For $Z_\star \geq 10^{-5}\,\Zsun$: AGB yield is from \citet{Karakas10, Doherty2014MNRAS}; SNIa yield is from \citet{Seitenzahl2013MNRAS}.}
}
\label{tab:modelsets}
\begin{tabular}{c|cc|cc|c|c}
\multicolumn{1}{c|}{} & \multicolumn{2}{c|}{$Z_\star < 10^{-5}\,\Zsun$} & \multicolumn{2}{c|}{$Z_\star \geq 10^{-5}\,\Zsun$} & \multicolumn{2}{c}{} \\
\multicolumn{1}{c}{} & \multicolumn{2}{|c|}{$0.7$--$300\,\Msun$} & \multicolumn{2}{c|}{$0.1$--$100\,\Msun$} & \multicolumn{2}{c}{} \\
\multicolumn{1}{c}{Model} & \multicolumn{2}{|c|}{IMF: \citet{Susa14}} & \multicolumn{2}{c|}{IMF: \citet{Chabrier03}} & \multicolumn{2}{c}{} \\
& CCSNe & AGB\&SNIa$^\dagger$ & CCSNe & AGB\&SNIa$^\sharp$ & SF & DC \\
\hline
N & \citet{Nomoto13} & yes & \citet{Nomoto13} & yes & continuous & no \\
$\mathrm{N}_\mathrm{int}$ & \citet{Nomoto13} & yes & \citet{Nomoto13} & yes & intermittent & no \\
L & \citet{Limongi2018ApJS} & yes & \citet{Limongi2018ApJS} & yes & continuous & yes \\
N\&N-10\% & \citet{Nandal2024AA} (10\%) & no & \citet{Nomoto13} & yes & intermittent & no \\
N\&N-10\%-DC & \citet{Nandal2024AA} (10\%) & no & \citet{Nomoto13} & yes & intermittent & yes \\
N\&N-CO & \citet{Nandal2024AA} (CO) & no & \citet{Nomoto13} & yes & intermittent & no \\
N\&N-CO-DC & \citet{Nandal2024AA} (CO) & no & \citet{Nomoto13} & yes & intermittent & yes \\
\hline
\end{tabular}
\end{center}
\end{table*}

\section{Result} \label{sec:result}


\subsection{Standard model: Model-N and Model-L} 
\label{subsec:res_onezone_standard}

\begin{figure*}[ht!]
\begin{center}
    \includegraphics[width=1.0\columnwidth]
    {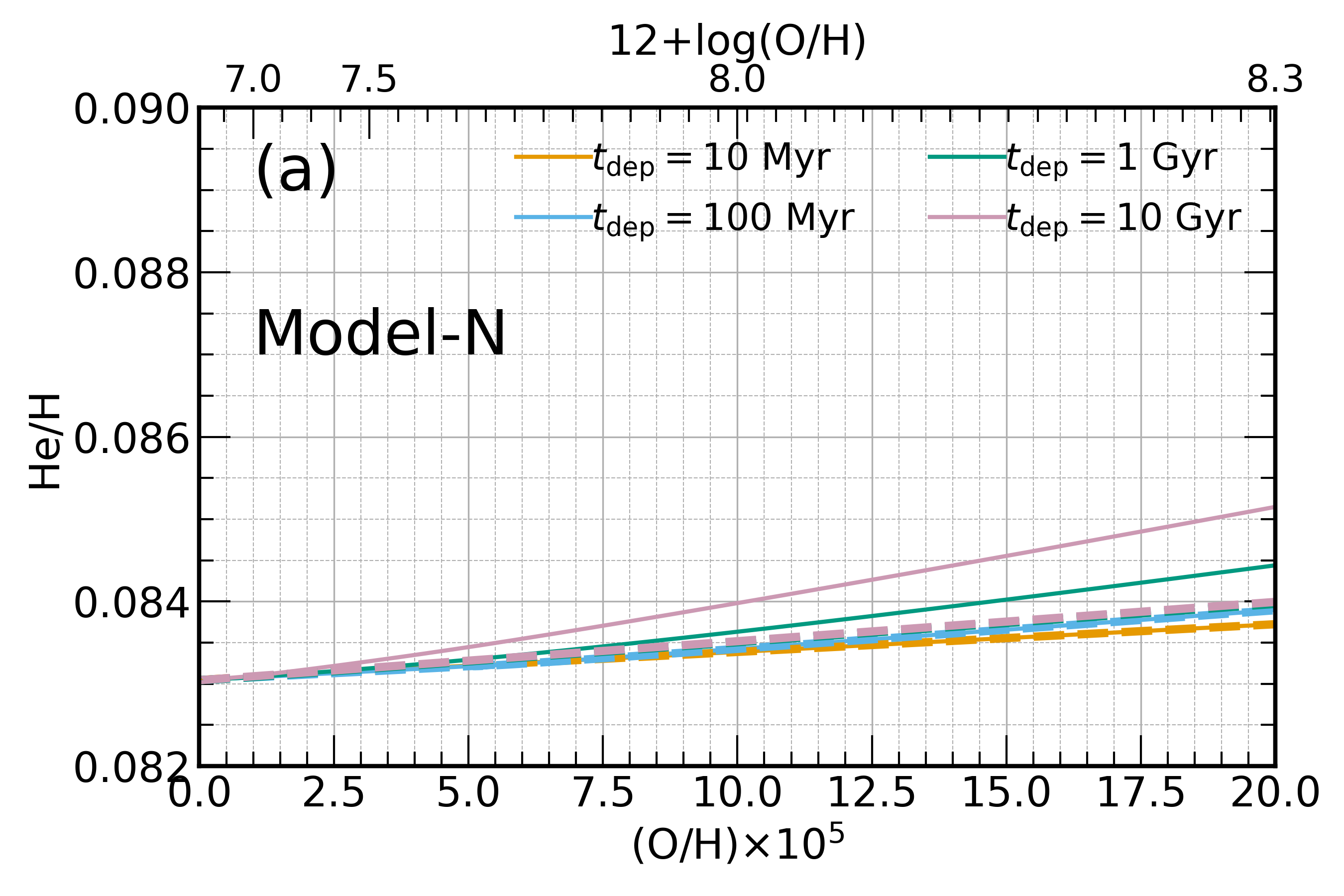}
    \includegraphics[width=1.0\columnwidth]
    {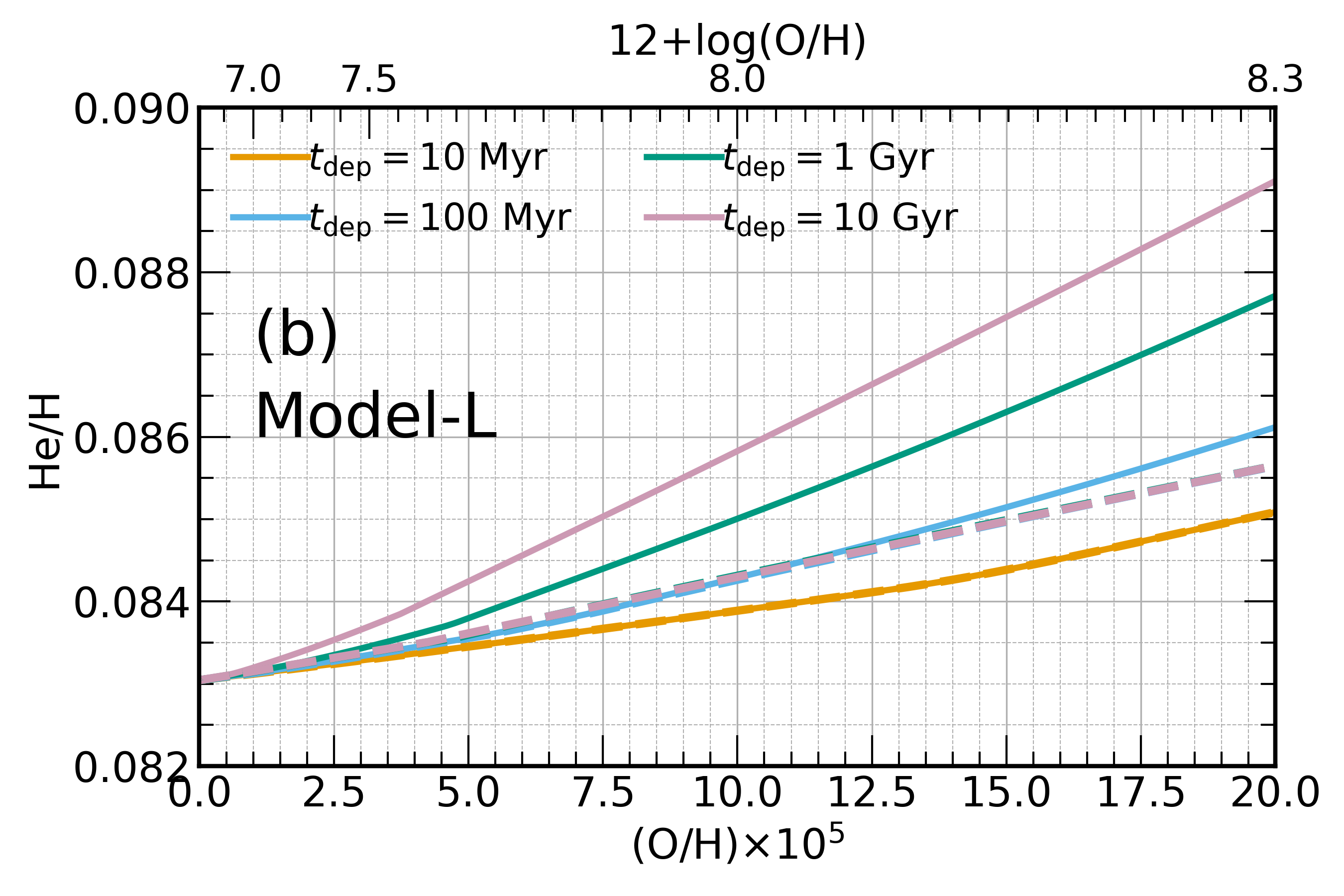}
\end{center}
\caption{
$(\mathrm{O/H})\times10^5$--$\mathrm{He/H}$ relationship derived from one-zone model calculations for Model-N ({\it panel (a)}) and Model-L ({\it panel (b)}). 
The orange, blue, green, and red lines represent $\ts=10$ Myr, 100 Myr, 1 Gyr, and 10 Gyr, respectively.  Solid lines include the contributions from  CCSN, SNIa, and AGB yields, while dashed lines represent models that consider only CCSNe.  The other model parameters are fixed at $\ti=1$ Gyr, $\fo=0.1$, and $f_{\mathrm{inf}}=0.1$.
In panel ({\it a}) (Model-N), the oxygen abundance reaches $\mathrm{O/H} \times 10^5 = 20$ after $4.3$ Myr, $53$ Myr, $400$ Myr, and $2.8$ Gyr for $\ts = 10^7$, $10^8$, $10^9$, and $10^{10}$ yr, respectively, illustrating the slower chemical evolution at longer $\ts$. 
In panel ({\it b}), for Model-L, the corresponding ages are $2.1 \times 10^7$\,yr, $1.3 \times 10^8$\,yr, $9.9 \times 10^8$\,yr, and $6.1 \times 10^9$\,yr for the same values of $\ts$. 
}
\label{fig:Nom_Lim_onezone}
\end{figure*}

We begin by presenting the chemical abundance trends from the standard one-zone model calculations.  Figure~\ref{fig:Nom_Lim_onezone} shows the relation between $(\mathrm{O/H})\times10^5$ and $\mathrm{He/H}$ for Model-N ({\it panel (a)}) and Model-L ({\it panel (b)}).
The orange, blue, green, and red lines represent models with gas depletion timescales of $\ts=10$\,Myr, 100\,Myr, 1\,Gyr, and 10\,Gyr, respectively. 
The solid lines indicate the results using the CCSN, SNIa, and AGB yields, while the dashed lines show results based solely on CCSN yields.

In both panels, the chemical enrichment is driven by CCSNe when the solid and dashed lines closely overlap. 
In contract, for models where the solid and dashed lines diverge, oxygen continues to be produced mainly by CCSNe, while the additional enrichment in helium is attributed to contributions from AGB stars.

\begin{figure*}[ht!]
\begin{center}
    \includegraphics[width=2.0\columnwidth]{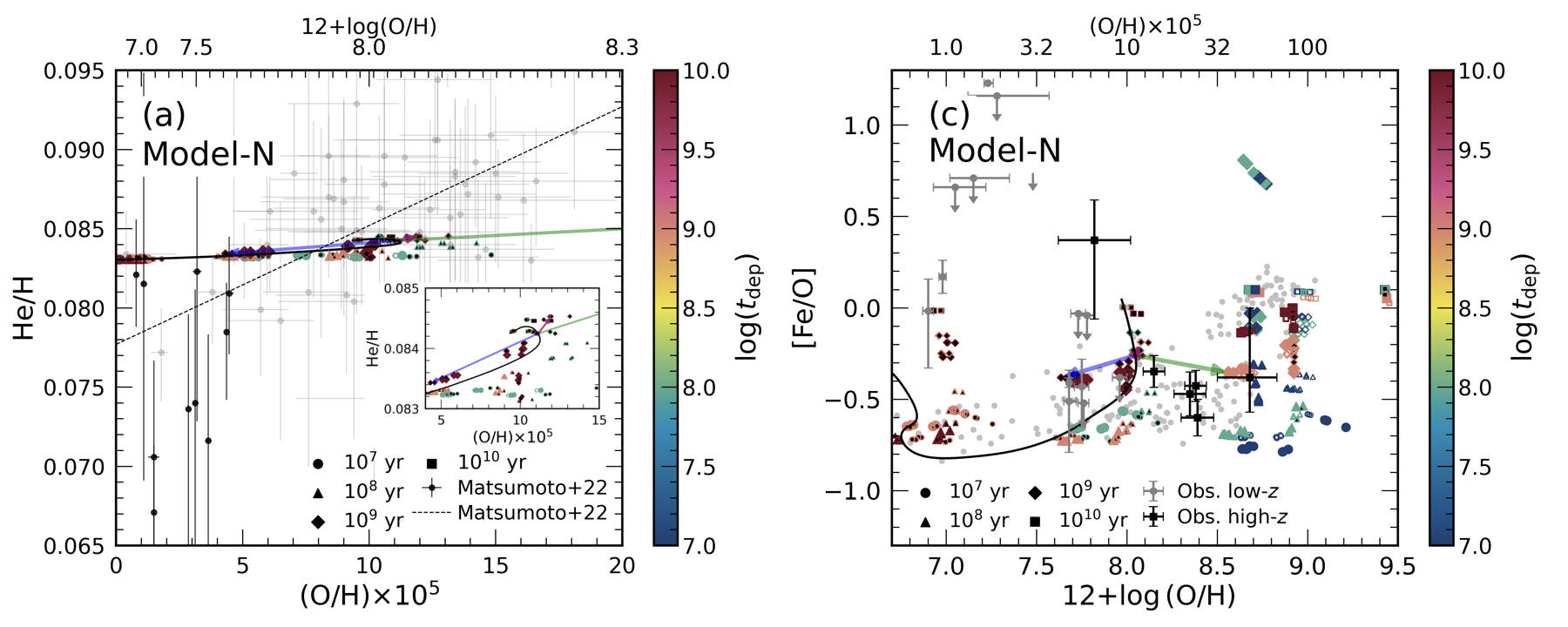}\\
    \includegraphics[width=2.0\columnwidth]{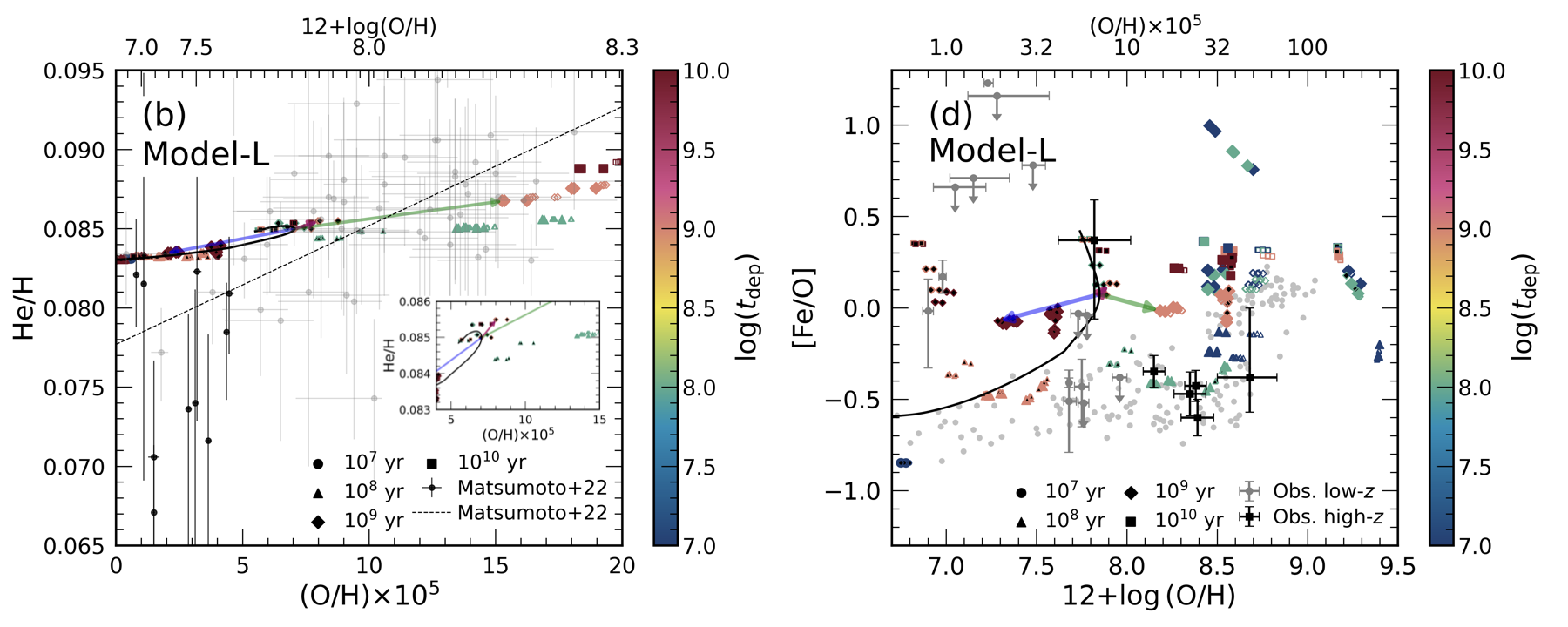}\\
\end{center}
\caption{{\it Panel (a):} Evolution of He/H as a function of metallicity in our Model-N. 
The different symbols represent the chemical abundance of galaxies at various evolutionary stages, with the color bar indicating $\ts$.
The style of the markers reflects the mass-loading factor $\fo$: markers with black-filled centers correspond to $\fo < 1$, fully filled markers indicate $\fo = 1$, and unfilled markers represent $\fo > 1$. Smaller-sized markers are used for the extreme cases of $\fo = 0.1$ and $\fo = 10$.
Black and gray points show observational results from \citet{Kojima2020ApJ, Izotov2012AandA, Thuan2005ApJS, Papaderos2008AandA, Izotov2019MNRAS, Nakajima2022ApJS} and \citet{Hsyu2020ApJ}, respectively (See also Table~\ref{tab:observed_gals_HeH}). 
The black dashed line represents the best-fit linear relation to the observed data from \citet{Matsumoto2022ApJ}. 
The solid black line shows the evolutionary track for the fiducial case of $(\ts[\mathrm{yr}], \ti[\mathrm{yr}], \fo, f_\mathrm{inf}) = (10^9, 10^{10}, 10, 0.01)$. 
Colored arrows indicate the effects of varying each parameter at $t = 10^9$ yr: the blue arrow corresponds to an increase in $\ts$ to $10^{10}$ yr, the red arrow to a decrease in $\ti$ to $10^9$ yr, the green arrow to a reduction in $\fo$ to 1, and the purple arrow to an increase in $f_\mathrm{inf}$ to 0.1.
The inset in {\it panel (a)} provides a zoomed-in view of the region $4 < \mathrm{O/H} \times 10^5 < 15$ and $0.083 < \mathrm{He/H} < 0.086$.
{\it Panel (b):} Same as panel ({\it a}), but using the Model-L yields from  \citet{Limongi2018ApJS}. 
{\it Panel (c):} Evolution of Fe/O as a function of metallicity. 
Data points with black error bars show the chemical abundances of high-$z$ galaxies \citep{Steidel16, Cullen2021MNRAS, Kashino2022ApJ, Harikane2020ApJ, Nakane2024ApJ}, while those with gray error bars represent low-$z$ galaxies \citep{Izotov2018MNRAS, Kojima2020ApJ, Kojima2021ApJ, Isobe2022ApJ} (See also Table~\ref{tab:observed_gals_FeO}).  Gray scatter points represent Milky Way stars \citep{Amarsi2019A&A}.
Galaxies with unusually high Fe/O at low metallicities, which are of particular interest in this study, are highlighted as solid black points.
{\it Panel (d):} Same as panel ({\it c}), but using the Model-L yields.}
\label{fig:onezone}
\end{figure*}

The results of the one-zone model calculations using the parameter sets listed in Table~\ref{tab:parameters} are shown in Figure~\ref{fig:onezone}, with outputs plotted at various galaxy ages.
Panels ({\it a} and {\it c}) employ the \citet{Nomoto13} yields (Model-N) to explore He/H and Fe/O ratios as a function of metallicity, respectively.  Panels ({\it b} and {\it d}) display the same quantities using the  \citet{Limongi2018ApJS} yields (Model-L). 

The scatter points represent galaxies at different evolutionary times: circles, triangles, crosses, squares, and pentagons correspond to ages of $10^6$, $10^7$, $10^8$, $10^9$, and $10^{10}$ yrs, respectively. 
The color of each marker indicates the gas depletion timescale, $\ts$, as defined in Appendix~\ref{sec:appendix_One-zone}. 
To visualize the effect of the outflow mass-loading factor, $\fo$, we vary the marker styles: markers with black-filled centers represent $\fo < 1$, fully filled markers correspond to $\fo = 1$, and unfilled markers indicate $\fo > 1$. For the extreme cases of $\fo = 0.1$ and $\fo = 10$, the marker size is reduced to enhance visibility.

A representative evolutionary track for the fiducial parameter set $(\ts[\mathrm{yr}], \ti[\mathrm{yr}], \fo, f_\mathrm{inf}) = (10^9, 10^{10}, 10, 0.01)$ is shown as a solid black line. 
The effects of varying each parameter individually are illustrated with colored arrows, which indicate the displacement of the track at $t = 10^9$\,yr. Specifically, the blue arrow corresponds to an increase in $\ts$ to $10^{10}$\,yr, the red to a decrease in $\ti$ to $10^9$\,yr, the green to a reduction in $\fo$ to 1, and the purple to an increase in $f_\mathrm{inf}$ to 0.1. Since the red and purple arrows nearly overlap, the red arrow is thickened and the purple arrow is thinned for clarity.

In panels ({\it a}) and ({\it b}), the black dots are the observations of EMPGs \citep{Kojima2020ApJ, Izotov2012AandA, Thuan2005ApJS, Papaderos2008AandA, Izotov2019MNRAS, Nakajima2022ApJS} summarized by \citet{Matsumoto2022ApJ}, and gray dots represent nearby dwarf galaxies from \citet{Hsyu2020ApJ}. The black dashed line indicates the best-fit linear relation derived by \citet{Matsumoto2022ApJ}. 

In panels ({\it c}) and ({\it d}), data points with black error bars show the chemical abundance of high-$z$ galaxies \citep{Steidel16, Cullen2021MNRAS, Kashino2022ApJ, Harikane2020ApJ, Nakane2024ApJ}, while those with gray error bars represent low-$z$ galaxies \citep{Izotov2018MNRAS, Kojima2020ApJ, Kojima2021ApJ, Isobe2022ApJ}. The gray scattered data points represent MW stars \citep{Amarsi2019A&A}. 

Here, the gas fraction is defined as 
\begin{equation}
    \label{eq:gas_fraction}
    \zeta_\mathrm{gas} = M_\mathrm{gas}/(M_\mathrm{gas}+M_\star),
\end{equation}
where $M_\mathrm{gas}$ and $M_\star$ are the gas and stellar mass, respectively, within the one-zone system. 
For the fiducial model shown by the black line in Figure~\ref{fig:onezone}, the gas fraction $\zeta_\mathrm{gas}$ evolves from 0.995 at $10^7$\,yr to 0.949 at $10^8$\,yr, 0.575 at $10^9$\,yr, and 0.076 at $10^{10}$\,yr.
In models with short $\ts$ or high $\fo$, gas is consumed or expelled more rapidly. Consequently, once the gas reservoir is depleted, the calculations are terminated. This leads to the absence of data points at later times (e.g., $10^9$–$10^{10}$\,yr) in some of the model tracks.

In panel ({\it a}), our one-zone model shows a discrepancy with certain EMPG observations; notably, none of our model results exhibit He/H ratios lower than $0.082$. This deviation mainly stems from the adoption of a higher primordial He abundance, as suggested by \citet{Planck16}, compared to the  He/H ratio at $(\mathrm{O/H})\times10^5=0$ determined by \citet{Matsumoto2022ApJ}. Additionally, the slope of our modeled He/H versus O/H relationship is shallower than that of the observed fitting line. This suggests that in our model, He enrichment from {\SNII} and AGB stars has a limited impact on altering the He/H abundance ratio, largely due to the predominance of primordial gas in the galaxy's composition as per our setup. 
Although He and oxygen are enriched by CCSNe, the He/H ratio in the CCSNe ejecta is at most around $\mathrm{He/H}\sim0.3$, which is only about 3.6 times higher than the primordial $\mathrm{He/H}\sim0.083$ (Appendix~\ref{sec:appendix_CELib}). 
However, the oxygen ejecta is extremely metal-rich ($12+\log(\mathrm{O/H})\sim11$), causing the one-zone box to be enriched with high-metallicity gas.
As a result, as seen in Figure~\ref{fig:Nom_Lim_onezone}, the slope becomes very shallow, producing data points with $\mathrm{He/H}\sim0.083$ across a wide range of $(\mathrm{O/H})\times10^5=0-15$, as seen in Figure~\ref{fig:onezone}.

Looking at the arrows, we can see that even when $\ti$ is shortened from $10^{10}$ years to $10^9$ years, if $\ts \leq \ti$, the changes are only around $\Delta (\mathrm{O/H}) \times 10^5 \sim 1$ and $\Delta \mathrm{He/H} \sim 0.0002$. Additionally, even when $f_{\mathrm{inf}}$ is increased by a factor of 10, the changes are of a similar magnitude to those when $\ti$ is changed. However, when $\fo$ is reduced by a factor of 10, no metal outflow occurs, and $\Delta (\mathrm{O/H}) \times 10^5 > 10$, resulting in a significant increase in $\mathrm{O/H}$.


In panel ({\it b}), similar to panel ({\it a}), we see that the He/H ratio does not fall below $0.082$. For values of $(\mathrm{O/H})\times10^5\gtrsim15$, our model achieves He/H ratios that are marginally lower than the fitting line of the observation established by \citet{Matsumoto2022ApJ}. 
As seen in Appendix~\ref{sec:appendix_CELib}, the $12+\log(\mathrm{O/H})$ released by CCSNe is $12+\log(\mathrm{O/H})=9-11$ for the \citet{Nomoto13} yield (Model-N), whereas for the \citet{Limongi2018ApJS} yield (Model-L), it only reaches a maximum of $12+\log(\mathrm{O/H})=10.5$. 
Additionally, for Model-L at $10^{6.8}$ years, stellar winds release gas with low metallicity, where $12+\log(\mathrm{O/H})<9$. In Model-L, 
the He/H released at this age is only about 0.3 times lower than that of Model-N, but since $12+\log(\mathrm{O/H})$ is more than 100 times lower, the gas can retain high He/H at low O/H. 
The trend of the black line and arrows is the same as in panel ({\it a}).
This yield model does not fully replicate the high He/H observations at low metal abundances noted by \citet{Hsyu2020ApJ}.


Some data points from our model calculations show a reduced offset from the fitting line from the observations \citep{Matsumoto2022ApJ} in the range where $(\mathrm{O/H}) \times 10^5 > 10$, compared to {\it panel (a)}.  
The difference is approximately $\Delta \mathrm{He/H} = 0.004$.
Those with the same $\ts$ are clustered within a range of $\Delta((\mathrm{O/H})\times10^5) = 5$ and $\Delta(\mathrm{He/H}) = 0.001$, and their ages are the same as well. 
By looking at the arrows, it becomes clear that this group shares the same $\fo$ and $\ts$ parameters, but has different $\ti$ and $f_\mathrm{inf}$ parameters. 
This indicates that the results in Figure~\ref{fig:Nom_Lim_onezone} show that changing $\ti$ and $f_\mathrm{inf}$ within the current parameter range only leads to variations within $\Delta((\mathrm{O/H})\times10^5) = 5$.
Although Figure~\ref{fig:Nom_Lim_onezone} suggests that longer $\ts$ leads to higher He/H, this trend is not seen in Figure~\ref{fig:onezone} because the arrows compare abundance ratios at fixed stellar ages, not at equivalent evolutionary stages.
When $\ts$ is larger, chemical evolution progresses more slowly compared to a shorter $\ts$, resulting in a lower He/H at the same stellar age.

In panel ({\it c}), for galaxies with $12+\log(\mathrm{O/H}) \sim 7.0$, our model shows lower [Fe/O] compared to observations at ages below 100 Myr, with an offset of $\Delta \mathrm{[Fe/O]} \sim 0.7$.  
This difference is primarily attributed to the central role of {\SNII} in the chemical evolution of these young galaxies in our model. Initially ($\sim10^7$ years), a high Fe/O ratio is observed due to metal enrichment by {\PopIII} stars (as detailed in Appendix~\ref{sec:appendix_CELib}), but this ratio is diminished over a period of approximately $10^{7.3}$ years by enrichment of $\alpha$ elements from conventional {\SNII}. 
Additionally, our results indicate that galaxies with $\ts = 10$ Gyr can exhibit relatively high Fe/O ratios ([Fe/O] $\sim -0.1$) in the age range of $1-10$ Gyr.
Our models with $\zeta_{\mathrm{gas}} < 0.95$ (see {\it panel (b)} in Appendix~\ref{sec:appendix_gasfrac}), are consistent with observations such as J1631$+$4426 ($\zeta_{\mathrm{gas}} \sim 0.91$) and J0811$+$4730 ($\zeta_{\mathrm{gas}} \sim 0.78$), which exhibit $12+\log(\mathrm{O/H}) \sim 7$ and [Fe/O] $\sim 0$ ( see Table~\ref{tab:observed_gals_FeO}).
Having higher Fe/O with lower star formation efficiency is also consistent with \citet{Vincenzo2014MNRAS}. 
To observe [Fe/O]$\sim0$ with $12+\log(\mathrm{O/H})<7.5$ with $\lesssim10^8$ yr, it may be necessary to use SN yields from PISNe or bright HNe before [Fe/O] decreases due to regular CCSNe \citep{Isobe2022ApJ, Nakane2024ApJ, Nakane2025arXiv}. Alternatively, a top-light IMF may need to be employed \citep{ Lee2009ApJ_dwarfgal, Yan2017AandA_IMF, Yan2020A&A, Mucciarelli2021NatAs, Nakane2025arXiv}. As shown in Appendix~\ref{sec:appendix_CELib}, relatively low-mass {\SNII} release high Fe/O ratios ([Fe/O]$\sim 0$).
However, EMPGs (J1631+4426, J0811+4730) exhibit sSFRs that are more than an order of magnitude higher than those of typical dwarf galaxies, suggesting that they may belong to a distinct population.

The trend of the black line and arrows for [Fe/O] is also the same as in panel ({\it a}).


Panel ({\it d}), also shows a galaxy with an age of $10^9$ yr with $[\mathrm{Fe/O}]\sim0.0$ and $12+\log(\mathrm{O/H})<7.5$, this chemical abundance close to the observed low-$z$ galaxies.
Furthermore, at $\log_{10} (\mathrm{O/H}) \sim 7.8$, there exist samples with [Fe/O] $\sim 0.4$ ($\ts = 1$ Gyr), which is comparable to the high-$z$ galaxy GN-z11 \citep{Isobe2023ApJ}. As indicated by the black line, this object may also have a low $\zeta_\mathrm{gas}$ similar to our sample.  
However, it is important to note that the age of our sample is $\gtrsim1$ Gyr, which does not match the observed value.

The Model-N is closer to the stellar data of the Milky Way, while the Model-L better matches the data of dwarf galaxies and high-$z$ galaxies with high [Fe/O]. 
The \citet{Limongi2018ApJS} yield Set R assumes that stars with masses above $25 M_\odot$ collapse directly into black holes, reducing the amount of metal ejection from massive stars.
This results in an outcome similar to that of a top-light IMF. 
Thus, the consistency of the Model-N with MW data and the similarity of the Model-L to dwarf galaxy data suggest that MW has a standard IMF, while dwarf galaxies may exhibit a top-light IMF \citep{Yan2020A&A, Mucciarelli2021NatAs}.

In summary, the use of \citet{Nomoto13} yields (Model-N) in our models successfully replicates the observed some dwarf galaxies' and the Galactic $12+\log$(O/H)-[Fe/H] for certain parameters. However, it falls short of accurately matching other aspects, such as the gas fraction, when compared to observations. 
In contrast, applying Model-L shows better agreement with the observed data of low-mass galaxies at low redshifts. This is because the yield from the set R of \citet{Limongi2018ApJS}, used in Model-L, assumes that massive stars with $>25\,\Msun$ undergo direct collapse into black holes, reducing metal ejection from massive stars. This reduction may explain the consistency with observations of dwarf galaxies with a top-light IMF.
\begin{deluxetable*}{ccccc}
\tablewidth{0pt}
\tablecaption{Observational data used for comparison with our model calculations.}
\tablehead{
\colhead{(1) ID} & \colhead{(2) He/H} & \colhead{(3) O/H$\times10^5$} & \colhead{(4) $\zeta_\mathrm{gas}$} & \colhead{(5) reference}
}
\startdata
J1631+4426 & $0.0617^{+0.0101}_{-0.0094}$ & $0.79\pm0.06$ & 0.91 & \citet{Kojima2020ApJ,Matsumoto2022ApJ,Xu2024ApJ} \\
J1016+3754 & $0.0778^{+0.0034}_{-0.0027}$ & $4.37\pm0.10$ & 0.7 & \citet{Izotov2012AandA,Matsumoto2022ApJ,Xu2024ApJ} \\
I Zw 18 NW & $0.0703^{+0.0032}_{-0.0035}$ & $1.49\pm0.04$ & 0.42 & \citet{Thuan2005ApJS,Matsumoto2022ApJ,Xu2024ApJ} \\
J1201+0211 & $0.0677^{+0.0078}_{-0.0063}$ & $3.12\pm0.11$ & - & \citet{Papaderos2008AandA,Matsumoto2022ApJ} \\
J1119+5130 & $0.0810^{+0.0043}_{-0.0040}$ & $3.20\pm0.17$ & - & \citet{Izotov2012AandA,Matsumoto2022ApJ} \\
J1234+3901 & $0.0804^{+0.0198}_{-0.0166}$ & $1.09 \pm 0.07$ & 0.98 & \citet{Izotov2019MNRAS,Matsumoto2022ApJ,Xu2024ApJ} \\
J0133+1342 & $0.0777^{+0.0065}_{-0.0056}$ & $3.64 \pm 0.11$ & - & \citet{Papaderos2008AandA,Matsumoto2022ApJ} \\
J0825+3532 & $0.0544^{+0.0142}_{-0.0048}$ & $2.86 \pm 0.08$ & - & \citet{Thuan2005ApJS,Matsumoto2022ApJ} \\
J0125+0759 & $0.0935^{+0.0096}_{-0.0055}$ & $4.47 \pm 0.19$ & 0.93 & \citet{Nakajima2022ApJS,Matsumoto2022ApJ,Xu2024ApJ} \\
J0935$-$0115 & $0.0688^{+0.0032}_{-0.0035}$ & $1.49 \pm 0.22$  & 0.96 & \citet{Nakajima2022ApJS,Matsumoto2022ApJ,Xu2024ApJ} \\
\hline
J0118+3512 \tablenotemark{a}  & $0.0792^{+0.0076}_{-0.0076}$ & $6.5^{+2.2}_{-1.3}$ & - & \citet{Hsyu2020ApJ} \\
\hline
GS$-$NDG$-$9422 & $0.104^{+0.007}_{-0.007}$ & $3.89^{+0.09}_{-0.90}$ & - & \citet{Cameron2024MNRAS, Yanagisawa2024ApJ} \\
RXCJ2248$-$ID & $0.166^{+0.018}_{-0.014}$ & $2.69^{+1.29}_{-0.50}$ & - & \citet{Topping2024MNRAS, Yanagisawa2024ApJ} \\
GLASS150008 & $0.142^{+0.066}_{-0.039}$ & $4.47^{+1.70}_{-0.75}$ & - & \citet{ Isobe2023ApJ, Yanagisawa2024ApJ} \\
\enddata
\tablecomments{Observed galaxies' (1) ID, (2) He/H, (3) O/H$\times10^5$, (4) $\zeta_\mathrm{gas}$, and (5) Reference. \label{tab:observed_gals_HeH}}
\tablenotetext{a}{One of their local galaxy samples is presented as an example.}
\end{deluxetable*}

Table~\ref{tab:observed_gals_HeH} summarizes the observational data plotted in {\it panels (a)} and {\it (b)} of Figure~\ref{fig:onezone}, while Table~\ref{tab:observed_gals_FeO} compiles the observational data plotted in {\it panels (c)} and {\it (d)}.  
The first 10 objects in Table~\ref{tab:observed_gals_HeH} correspond to the galaxies compiled by \citet{Matsumoto2022ApJ}.  
In Table~\ref{tab:observed_gals_FeO}, the first 13 samples are classified as low-$z$ galaxies, while the latter 10 samples are categorized as high-$z$ galaxies.

\begin{longrotatetable}
\tabletypesize{\footnotesize}
\centering
\begin{deluxetable}{ccccccccc}
\tablewidth{0pt}
\tablecaption{Observational data used for comparison with our model calculations.}
\tablehead{
\colhead{(1) ID} & \colhead{(2) $z$} & \colhead{(3) $\log(M_\star/\Msun)$} & \colhead{(4) $\log$ SFR [$\Msun/\mathrm{yr}$]} & \colhead{(5) 12 + log(O/H)} & \colhead{(6) log(N/O)} & \colhead{(7) [Fe/O]} & \colhead{(8) $\zeta_\mathrm{gas}$} & \colhead{(9) reference}
}
\startdata
J0156$-$0421 & 0.04907 & 5.7 & $-1.6$ & $7.48^{+0.07}_{-0.06}$ & $-1.22^{+0.10}_{-0.11}$ & $< 0.78$ \tablenotemark{a} & - & {\it A}, {\it B} \\
J0159$-$0622 & 0.00852 & 5.4 &  & $7.68^{+0.04}_{-0.03}$  & $-1.47^{+0.03}_{-0.03}$ & $-0.51^{+0.17}_{-0.28}$ \tablenotemark{a} & - & {\it B} \\
J0210$-$0124 &  0.01172 & 5.3 &  & $7.76^{+0.03}_{-0.03}$  & $-1.72^{+0.03}_{-0.03}$ & $< -0.52$ \tablenotemark{a} & - & {\it B} \\
J0214$-$0243 & 0.02860 & 6.1 &  & $7.96^{+0.04}_{-0.04}$ & $-1.51^{+0.03}_{-0.03}$ & $< -0.38$ \tablenotemark{a} & - & {\it B} \\
J0226$-$0517 & 0.04386 & 6.4 & $-1.2$ & $7.78^{+0.04}_{-0.04}$ & $-1.51^{+0.06}_{-0.07}$ & $< -0.04$ \tablenotemark{a} & - & {\it A}, {\it B} \\
J0232$-$0248 &  0.04336 & 5.9 &  & $7.73^{+0.04}_{-0.04}$  & $-1.79^{+0.10}_{-0.12}$ & $< -0.03$ \tablenotemark{a} & - & {\it B} \\
J1608+4337 & 0.02896 & 6.0 &  & $7.75^{+0.04}_{-0.04}$ & $-1.47^{+0.02}_{-0.03}$ & $-0.43^{+0.15}_{-0.22}$ \tablenotemark{a} & - & {\it B} \\
J2236+0444 & 0.02870 & 5.2 & $-1.7$ & $7.05^{+0.17}_{-0.12}$ & $< -1.26$  & $< 0.66$ \tablenotemark{a} & - & {\it A}, {\it B} \\
J2321+0125 &  0.05639 & 5.4 & $-2.0$ & $7.28^{+0.29}_{-0.16}$  & $< -0.68$ & $< 1.16$ \tablenotemark{a} & - & {\it A}, {\it B} \\
J2355+0200 &  0.01231 & 4.8 & $-1.0$ & $7.15^{+0.20}_{-0.13}$  & $< -1.13$ & $< 0.71$ \tablenotemark{a} & - & {\it A}, {\it B} \\
J1631+4426 & 0.03125 & 5.89 & $-1.28$ & $6.90^{+0.03}_{-0.03}$ & $<-1.710$ & $-0.016^{+0.174}_{-0.313}$ \tablenotemark{a} & 0.91 & {\it C}, {\it D} \\
J2115$-$1734 & 0.02296 & 6.56 & $0.27$ & $7.68^{+0.01}_{-0.01}$ & $-1.518^{+0.009}_{-0.011}$ & $-0.41^{+0.026}_{-0.026}$ \tablenotemark{a} & 0.97 & {\it C}, {\it D} \\
J0811+4730 & 0.04444 & 6.29 & $0.48$ & $6.979^{+0.019}_{-0.019}$ & $-1.535^{+0.044}_{-0.044}$ & $0.17^{+0.092}_{-0.092}$ \tablenotemark{a} & 0.78 & {\it E}, {\it D} \\
\hline
GS$-$NDG$-$9422 & 5.94 & 7.8 \tablenotemark{b} & 8.2 & $7.59^{+0.01}_{-0.01}$ & $<-0.85$ & - & - & {\it F}, {\it G} \\
RXCJ2248$-$ID & 6.11 & 8.05 & 1.8 & $7.43^{+0.17}_{-0.09}$ & $-0.39^{+0.10}_{-0.08}$ & - & - &  {\it H} \\
GLASS150008 & 6.23 & 8.39 & 10 & $7.65^{+0.14}_{-0.08}$ & $-0.40^{+0.05}_{-0.07}$ & - & - & {\it I}, {\it J} \\
CEERS$\_$01019 & 8.68 & 9.5 & 1.6 & $7.94^{+0.46}_{-0.31}$ & $>0.28$ & - & - & {\it K}, {\it J} \\
GN$-$z11 & 10.6 & 9.1 & 21 & $8.00^{+0.76}_{-0.46}$ & $> -0.36$ & $0.37^{+0.22}_{-0.43}$ & - & {\it L}, {\it J}, {\it M} \\
low-$M_\star$ & 3.4 & 9.09 & - & $8.15^{+0.06}_{-0.06}$ & - & $-0.348^{+0.087}_{-0.087}$ & - & {\it N} \\
high-$M_\star$ & 3.4 & 9.81 & - & $8.38^{+0.06}_{-0.06}$ & - & $-0.4266^{+0.085}_{-0.085}$ & - & {\it N} \\
- & $\sim6$ & - & - & $8.7^{+0.2}_{-0.2}$ & - & $-0.38^{+0.38}_{-0.19}$ & - &  {\it O}\\
KBSS-LM1 & $2.4$ & 9.8 & 1.43 & $8.39^{+0.09}_{-0.09}$ & - & $-0.6^{+0.1}_{-0.1}$ & - &  {\it P}\\
1336 gals. & $1.6<z<3.0$ & 9.97 & 2.1 & $8.35^{+0.09}_{-0.09}$ & - & $-0.47^{+0.12}_{-0.12}$ & - & {\it Q} \\
\enddata
\tablecomments{Observed galaxies' (1) ID or sample name, (2) $z$, (3) $\log(M_\star/\Msun)$, (4) $\log$ SFR [$\Msun/\mathrm{yr}$], (5) 12 + log(O/H), (6) log(N/O), (7) [Fe/O], and (8) $\zeta_\mathrm{gas}$. (9) Reference {\it A} is \citet{Isobe2021ApJ}, {\it B} is \citet{Isobe2022ApJ}, {\it C} is \citet{Kojima2021ApJ}, {\it D} is \citet{Xu2024ApJ}, {\it E} is \citet{Izotov2018MNRAS_J0811+4730}, {\it F} is \citet{Terp2024AandA}, {\it G} is \citet{Cameron2024MNRAS}, {\it H} is \citet{Topping2024MNRAS}, {\it I} is \citet{Jones2023ApJ}, {\it J} is \citet{Isobe2023ApJ}, {\it K} is \citet{Larson2023ApJ}, {\it L} is \citet{Tacchella2023ApJ}, {\it M} is \citet{Nakane2024ApJ}, {\it N} is \citet{Cullen2021MNRAS}, {\it O} is \citet{Harikane2020ApJ}, {\it P} is \citet{Steidel16}, and {\it Q} is \citet{Kashino2022ApJ}.}
\tablenotetext{a}{We use the solar abundance of $\log$(Fe/O)=$-1.23$ \citep{Asplund2021AandA}}
\tablenotetext{b}{While \citet{Terp2024AandA} estimate the stellar mass of GS$-$NDG$-$9422 assuming a stellar continuum and report $\log (M_\star/\Msun) = 7.80\pm0.01$, \citet{Cameron2024MNRAS} assume that the nebular continuum dominates and do not provide a stellar mass estimate.\label{tab:observed_gals_FeO}}
\end{deluxetable}
\end{longrotatetable}

\begin{figure*}[ht!]
\begin{center}
    \includegraphics[width=1.0\columnwidth]
    {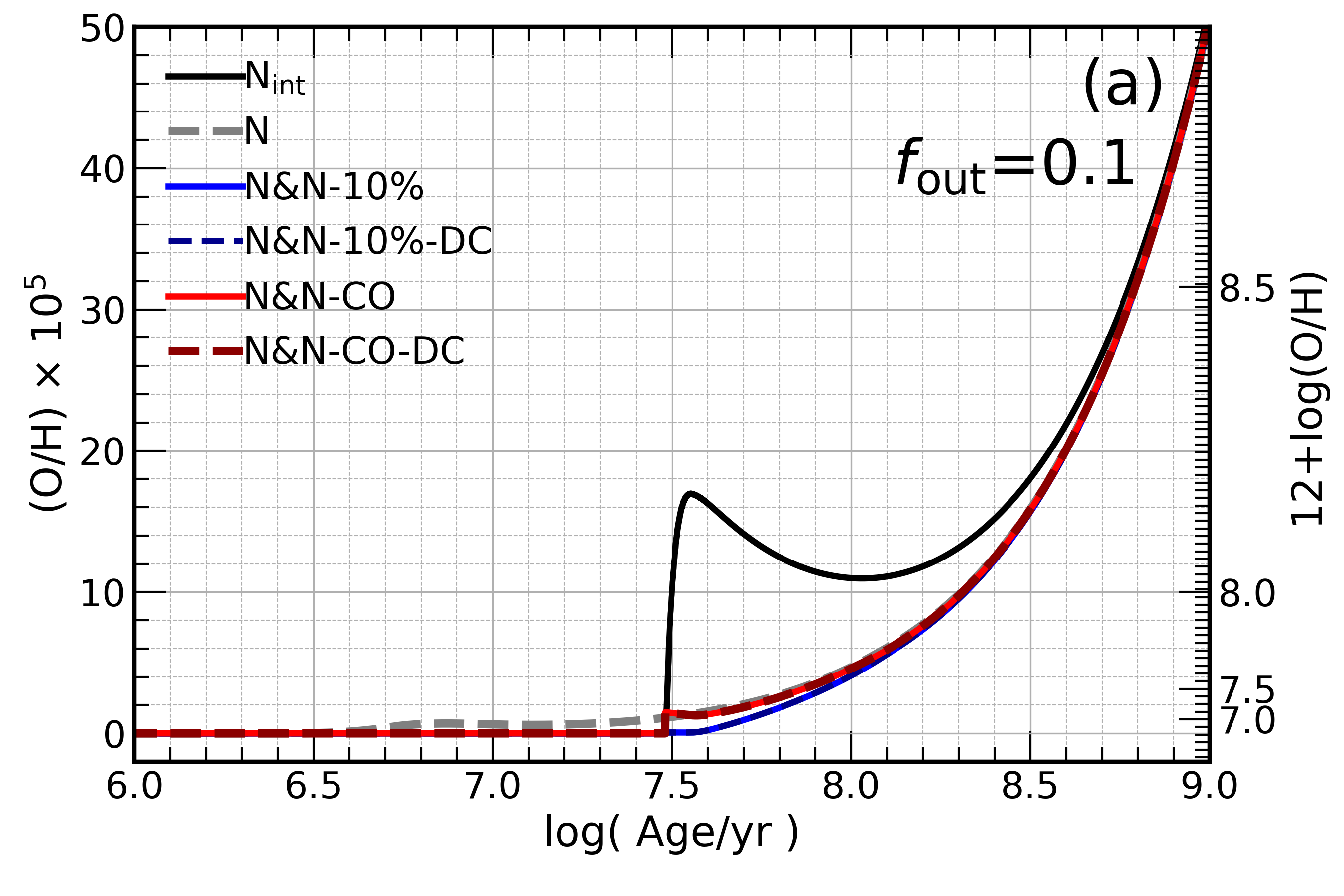}
    \includegraphics[width=1.0\columnwidth]
    {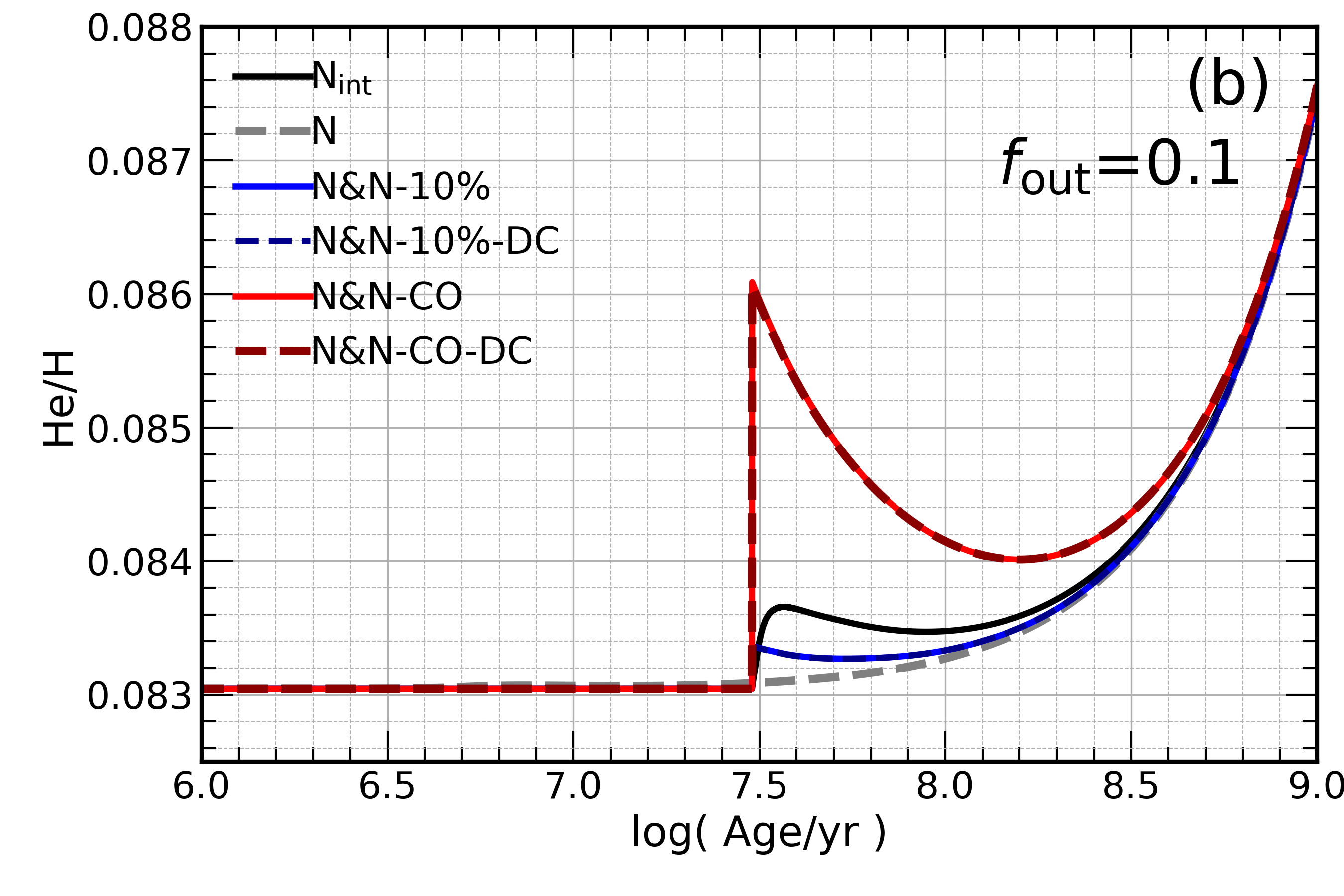}\\
    \includegraphics[width=1.0\columnwidth]
    {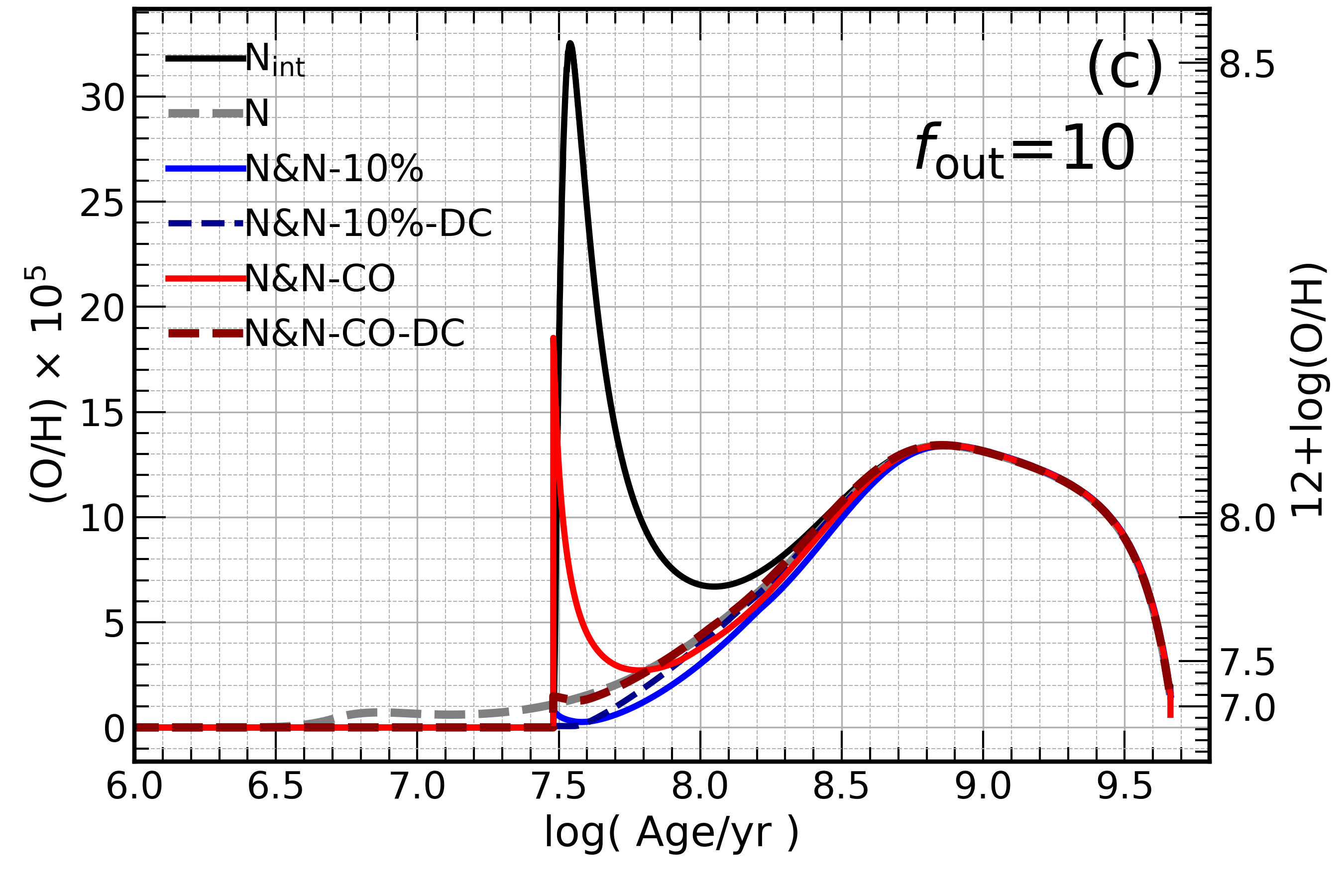}
    \includegraphics[width=1.0\columnwidth]
    {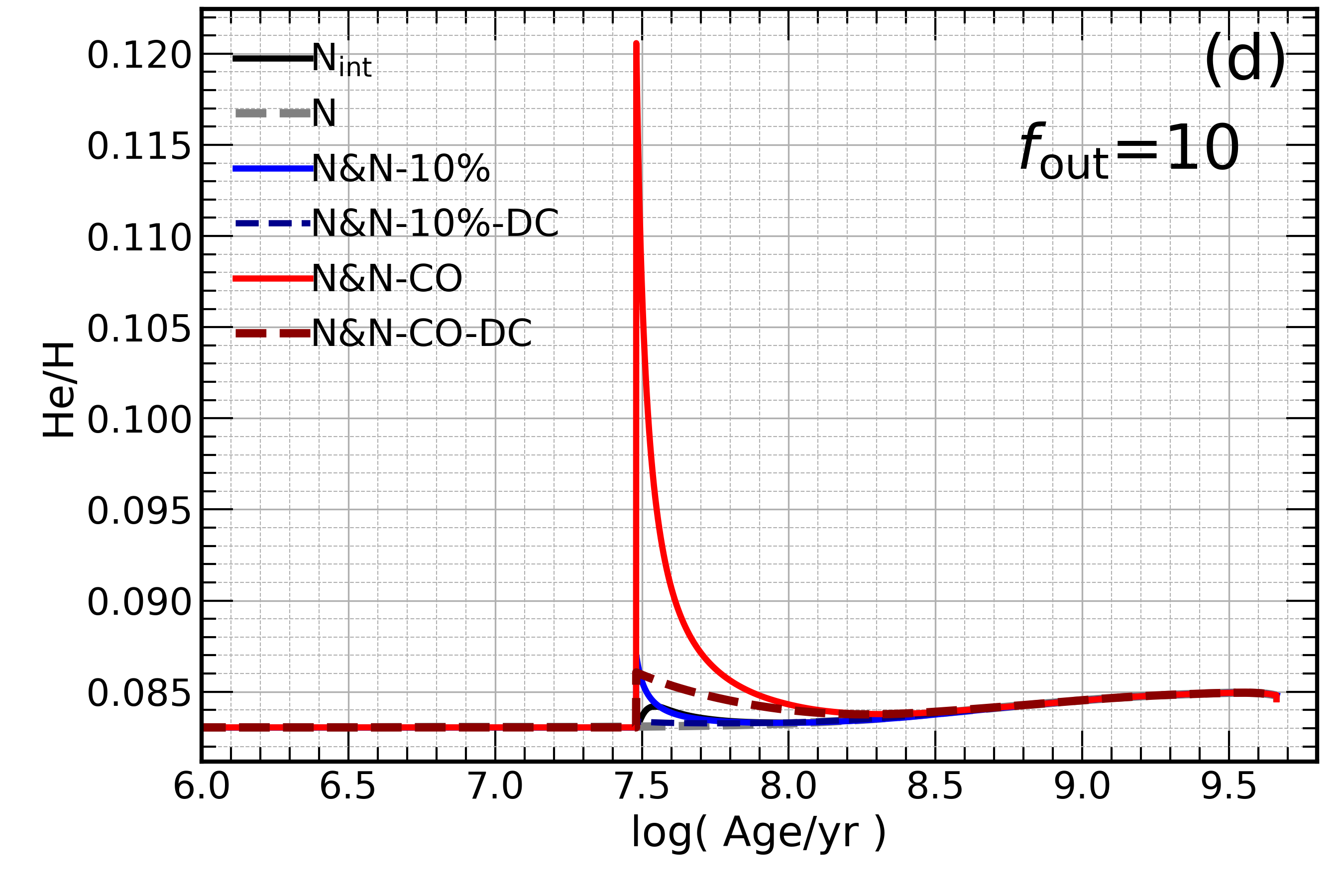}
\end{center}
\caption{
 {\it Panel (a)}: Time evolution of $\mathrm{O/H}\times10^5$ for 
 Model-N$_\mathrm{int}$ (black solid line), Model-N (gray dashed), Model-N\&N-$10\%$ (blue solid), Model-N\&N-$10\%$-DC (dark blue dashed), Model-N\&N-CO (red solid), and Model-N\&N-CO-DC (dark red-dashed).
 The gas accretion begins at $t=0$. 
Here, $\ts=10^9 \,\mathrm{yr}$, $\ti=10^9 \,\mathrm{yr}$, $\fo=0.1$, and $f_\mathrm{inf}=0.1$ are used (same as Figure~\ref{fig:Nom_Lim_onezone}).
{\it Panel (b)}: Time evolution of He/H for the same models as in {\it panel (a)}.
{\it Panel (c)}: Same as {\it panel (a)}, but with $\fo = 10$.
{\it Panel (d)}: Same as {\it panel (b)}, but with $\fo = 10$.
\label{fig:time_evolution_onezoneSMS}}
\end{figure*}

\subsection{Enrichment by Supermassive Stars}\label{subsec:SMSEnrichment}
\subsubsection{\rm He/H vs. O/H}\label{subsubsec:He/H}
As seen in Figs.~\ref{fig:Nom_Lim_onezone} and \ref{fig:onezone}, it is difficult to achieve a high He/H (He/H $>$ 0.1) like that of \citet{Yanagisawa2024ApJ} with low metallicity ($\mathrm{O/H}\times10^5 < 5$) in our standard model, even when using Limongi yield (Model-L), which considers stellar rotation. Therefore, in the following we also present the results of models that assume that {\PopIII} stars are SMS and using an intermittent star formation model. 

Figure~\ref{fig:time_evolution_onezoneSMS} shows the time evolution of O/H ({\it panel (a), (c)}) and He/H ({\it panel (b), (d)}) for models using SMS yield and for a star formation model with intermittent star formation. 
The parameters were set as $\ts=10^9 \,\mathrm{yr}$, $\ti=10^9 \,\mathrm{yr}$, $\fo=0.1$, and $f_\mathrm{inf}=0.1$ for {\it panels (a), (b)}, and $\ts=10^9 \,\mathrm{yr}$, $\ti=10^9 \,\mathrm{yr}$, $\fo=10$, and $f_\mathrm{inf}=0.1$.

In {\it panel (a)} ({\it (c)}), Model-N\&N-CO reaches $\mathrm{O/H} \times 10^5 \sim 2$ ($\sim 19$) at $10^{7.5}$ years, which is lower than that of Model-N$_\mathrm{int}$. 
In contrast, in {\it panel (b)} ({\it (d)}), Model-N\&N-CO reaches $\mathrm{He/H} = 0.086$ ($\mathrm{He/H} = 0.12$) at $10^{7.5}$ years, which is higher than Model-N$_\mathrm{int}$. 
As shown in the Appendix~\ref{sec:appendix_CELib}, this is because the SMS yield from \citet{Nandal2024AA} has higher He/H and lower O/H compared to the Nomoto yield.
Here, Model-N$_\mathrm{int}$ shows higher O/H compared to Model-N due to the presence of strong outflows at $10^{7.5}$ years (see Appendix~\ref{sec:appendix_outflow}).

The comparison between Model-N\&N-CO-DC and Model-N\&N-CO shows that stronger outflows (i.e., $\fo = 10$, without considering DC) result in higher He/H and O/H $\times 10^5$.
For $\fo = 10$, the gas fraction becomes lower compared to the case with $\fo = 0.1$, and 12+log(O/H) approaches the yield values of SNe Ia and AGB stars (see Figure~\ref{fig:Appen_CElib_OH_HeH_FeO}, {\it panel (a)}).

The Model-N\&N-10\% shows lower O/H and He/H compared to Model-N\&N-CO. 
This is because the ejecta from the case where $10\%$ of the SMS mass is released have lower O/H and He/H compared to the ejecta from the case where all outer layers up to the CO core are ejected (see also Appendix~\ref{sec:appendix_CELib}), and the ejecta mass in the $10\%$ case is eight times smaller than that in Model-N\&N-CO.

\begin{figure*}[ht!]
\begin{center}
    \includegraphics[width=2\columnwidth]{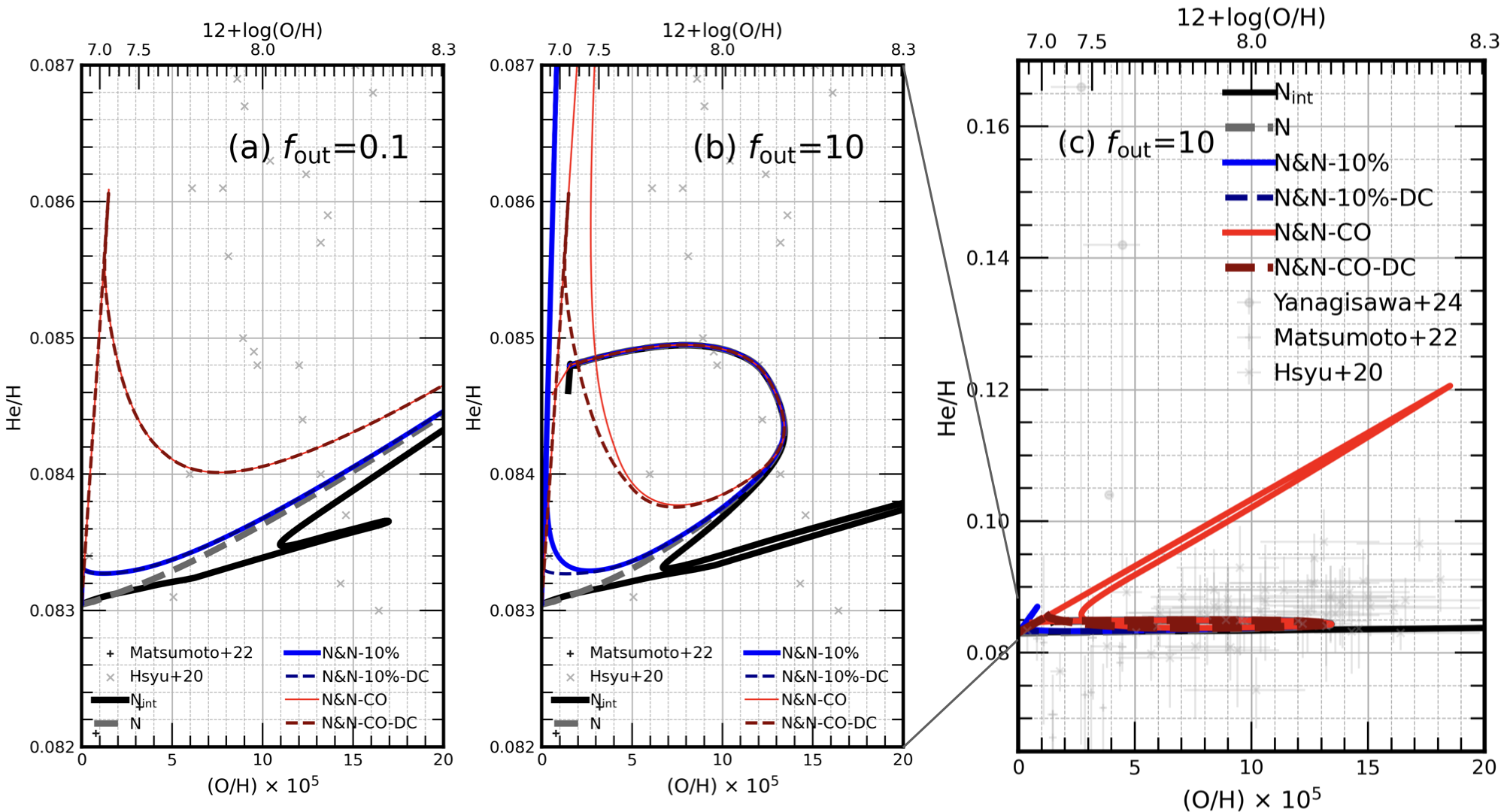}
\end{center}
\caption{
The same as Figure~\ref{fig:Nom_Lim_onezone}, but using the SMS yield of \citet{Nandal2024AA} (Model-N\&N series).
{\it Panel (a)} shows $\ti=1$ Gyr, $\ts=1$ Gyr, $\fo=0.1$, and $f_{\mathrm{inf}}=0.1$, as in Figure~\ref{fig:time_evolution_onezoneSMS}. {\it Panel (b)} plots the same parameters except with $\fo = 10$ for comparison. {\it Panel (c)} displays an expanded $y$-axis version of {\it panel (b)} to compare with He/H in high-$z$ galaxies.
The line types are also the same as in Figure~\ref{fig:time_evolution_onezoneSMS}.
The data points with gray error bars represent the observed He/H and $\mathrm{O/H} \times 10^5$ values for local galaxies \citep{Hsyu2020ApJ, Matsumoto2022ApJ}, while the data points with black error bars represent the observed He/H and $\mathrm{O/H} \times 10^5$ values for high-redshift galaxies \citep{Yanagisawa2024ApJ}.}
\label{fig:SMS_onezone}
\end{figure*}

Figure~\ref{fig:SMS_onezone} shows the relation between $(\mathrm{O}/\mathrm{H}) \times 10^5$ and $\mathrm{He}/\mathrm{H}$. 
{\it Panel (a)} shows $\ti=1$ Gyr, $\ts=1$ Gyr, $\fo=0.1$, and $f_{\mathrm{inf}}=0.1$, as in Figure~\ref{fig:time_evolution_onezoneSMS}. {\it Panel (b)} plots the same parameters except with $\fo = 10$ for comparison. {\it Panel (c)} displays an expanded $y$-axis version of {\it panel (b)} to compare with He/H in high-$z$ galaxies.
The line types are the same as in Figure~\ref{fig:time_evolution_onezoneSMS}.
Data points with gray error bars show the observed galaxies 
by \citet{Hsyu2020ApJ, Matsumoto2022ApJ, Yanagisawa2024ApJ}.

In the case of $\fo = 0.1$, both Model-N\&N-CO and Model-N\&N-CO-DC show higher He/H than Model-N when O/H $\times 10^5 < 20$, however, they do not reach the observed value of He/H $> 0.1$ in high-$z$ galaxies.  
For $\fo = 10$, He/H exceeds 0.1 in Model-N\&N-CO; however, O/H $\times 10^5$ is four times higher than the observed value for high-$z$ galaxies (O/H $\times 10^5 < 5$).  

Additionally, {\it panel (b)} shows that the He/H–O/H $\times 10^5$ relation due to enrichment by SMS in Model-N\&N-10\% is steeper compared to Model-N\&N-CO. 

\subsubsection{\rm N/O vs. O/H}
\begin{figure}[ht!]
\begin{center} 
    \includegraphics[width=1.0\columnwidth]
    {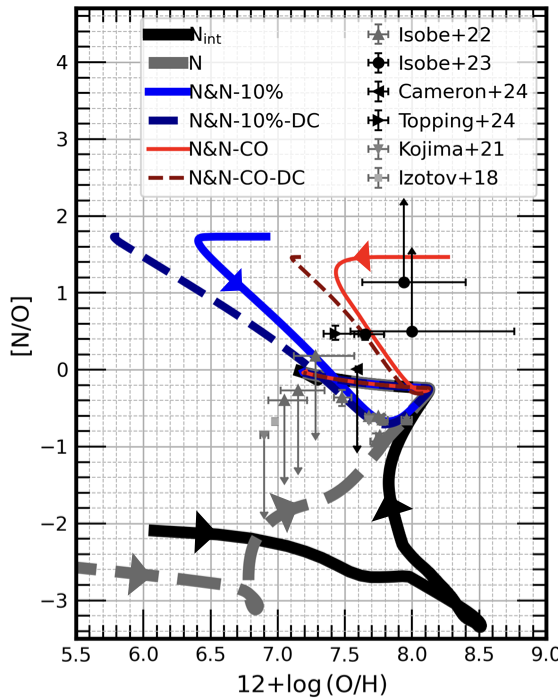}
\end{center}
\caption{
The evolutionary track of [N/O] and $12+\log(\mathrm{O/H})$ using the same yield set, star formation history, and parameters as those used in Figure~\ref{fig:SMS_onezone}.
However, here we plot from $\zeta_\mathrm{gas}=1$ to $\zeta_\mathrm{gas}=0$ with small arrows indicating the direction of time evolution. 
Here, $\ts=10^9 \,\mathrm{yr}$, $\ti=10^9 \,\mathrm{yr}$, $\fo=10$, and $f_\mathrm{inf}=0.1$ were also set.
Data points with gray error bars show the observed galaxies 
by \citet{Izotov2018MNRAS_J0811+4730,Kojima2021ApJ, Isobe2022ApJ, Isobe2023ApJ, Cameron2024MNRAS, Topping2024MNRAS}.
\label{fig:SMS_NO_onezone}
}
\end{figure}

In Figure~\ref{fig:SMS_NO_onezone}, we plot the evolution in [N/O] vs. $12+\log$(O/H) using the same yield set,  star formation history, and line types as in Figure~\ref{fig:SMS_onezone}. 
Because N/O is undefined when O/H = 0, and we assume that SN-ejected metals are instantaneously mixed with the gas, the plotted tracks begin only after both nitrogen and oxygen have been injected into the system by {\PopIII} stars.


We can see that SMS can raise N/O to [N/O]$ = 1.8$ (in Model-N\&N-10\%) and 1.5 (in Model-N\&N-CO), while still maintaining a low metallicity of $12+\log$(O/H) $< 8.3$.
As regular CCSNe contribute to further chemical enrichment, the initially high [N/O] gradually declines and eventually matches the levels predicted by Model-N around $12+\log$(O/H)$ = 8.05$, with [N/O]$ = -0.3$.
Our model also appears to reproduce the high N/O ratios observed in high-$z$ galaxies through the contribution of SMS.

\section{Discussion} \label{sec:discussion}

\subsection{Impact of SMS and intermittent star formation model}\label{subsec:discussion_SMS_starburst}

Metal enrichment by SMS is also being considered as a solution to the N/O abundance. The mass range of $\Msms$ used in Figure~\ref{fig:SMS_onezone} falls within the range that reproduces the N/O ratios observed in high-$z$ galaxies by JWST, as identified by \citet{Nandal2024AA}, and appears to be effective in achieving high He/H at young ages with $\mathrm{O/H}\times10^5 < 20$.
Furthermore, \citet{Nandal2025arXiv} suggests that SMSs can reproduce not only the high N/O ratio of GS 3073 at $z = 5.55$, but also its C/O and Ne/O ratios.  
These findings highlight the potential of SMSs to account for elevated He/H, N/O, C/O, and Ne/O ratios observed in some galaxies. Nonetheless, significant uncertainties remain regarding the physical properties, formation mechanisms, and feedback effects of SMSs.


Additionally, \citet{Kobayashi_Ferrara2024ApJ} pointed out that in the case of an intermittent star formation history, a high N/O can be achieved at young ages using a one-zone model.  
Both models demonstrate a similar trend in which metallicity decreases during the quiescent phase of star formation and then increases again through chemical evolution. However, the duration of this quiescent phase differs substantially between the models — approximately 2 Myr in our model $N_{\text{int}}$, compared to 100 Myr in the scenario proposed by \citet{Kobayashi_Ferrara2024ApJ}. 
This significant difference in timescales highlights that the two models are fundamentally distinct.
This suggests that an intermittent star formation history, as shown by cosmological zoom-in hydrodynamic simulations \citep{Yajima2017ApJ, Arata2019MNRAS, Arata2020MNRAS, Hirai2024ApJ}, is important for the chemical abundance ratios of high-$z$ galaxies.

\subsection{How to Estimate the Primordial He/H Ratio}\label{subsec:How_to_estimate}

The insights drawn from Figure~\ref{fig:onezone} suggest that, to determine the primordial He/H from observational data, it is necessary to accurately measure the He/H ratio of galaxies before the gas fraction decreases to 0.6. As indicated by the black line in Figure~\ref{fig:onezone}, the He/H ratio increases sharply as the gas fraction decreases to 0.6. Furthermore, in the current model, the O/H$\times10^5$ does not increase significantly because CCSNe do not effectively release oxygen at that age. By excluding galaxies with low gas fractions ($\zeta_\mathrm{gas}<0.6$) that exhibit high He/H ratios ($>0.09$), 
it may be possible to obtain a more robust relation to estimate primordial He abundance. 




To handle more realistic gas inflows, star formation histories, and outflows beyond the one-zone model, it is essential to perform simulations of EMPGs and dwarf galaxies with high mass resolution. 
As shown in Figure~\ref{fig:SMS_onezone}, the He/H ratios observed in high-$z$ galaxies by \citet{Yanagisawa2024ApJ} can potentially be reproduced by the yields from SMSs.  
In particular, while the He/H value varies depending on the ratio between the gas mass in the box and the SMS mass, the slope remains unchanged. This suggests that using the yields from Model-N\&N-10\% might allow for the reproduction of both He/H and O/H $\times 10^5$.  
Applying the yields from Model-N\&N-10\% in cosmological hydrodynamic simulations to solve realistic gas dynamics is left for future work.
In particular, such a simulation may answer the question of whether EMPGs are local analogs of the first galaxies. 

An obvious future task is to expand the dataset of He/H and O/H observations in galaxies with low metallicity and compare the fitting curves across different specific SFRs.
This requires deep spectroscopic observations of many dwarf galaxies, including EMPGs. Information on the He/H--O/H relation of high-$z$ galaxies, as observed by JWST, is also important and will be the subject of future work.

\section{Summary} \label{sec:summary}

We investigate the chemical evolution of EMPGs 
employing the one-zone box model with different yield models. 
The findings from our model indicate that galaxies with long gas-depletion timescales achieve high He/H at low metallicity, similar to the observed data, when using the \citet{Limongi2018ApJS} yield (Model-L) which includes metal enrichment from the WR star (See Figure~\ref{fig:Nom_Lim_onezone}). However, in terms of $(\mathrm{O/H}) \times 10^5$, our samples' He/H from the fitting line by \citet{Matsumoto2022ApJ} was smaller by $\Delta \mathrm{He/H} = 0.003$.
Moreover, Model-L successfully reproduces a high Fe/O ratio ([Fe/O] $\sim 0.0$) under low-metallicity ($12+\log(\mathrm{O/H}) < 8.0$). This result is consistent with the observed EMPGs and the high-$z$ galaxy GN-z11.

Using SMS yields (Model-N\&N series) can further help explain galaxies with metallicities of $(\mathrm{O/H}) \times 10^5 < 20$ and $\mathrm{He/H} > 0.085$ at young ages ($< 10^8$ yrs).
Additionally, our Model-N\&N-CO can achieve $\mathrm{He/H} > 0.12$, comparable to the high-$z$ galaxies found by JWST. 
These SMS yield models also show high $[\mathrm{N/O}] > 0.3$, as observed by JWST in high-$z$ galaxies.

Finally, we discuss future prospects.
To more accurately reproduce the observed chemical abundance in young, low-metallicity galaxies such as EMPGs and first galaxies, it is necessary to perform high-resolution cosmological hydrodynamic simulations that can realistically model baryon cycling down to $z=0$.
These simulations should focus on the formation and evolution of EMPGs and dwarf galaxies, capturing starbursts and the effects of different stellar yield models, including those of rotating massive stars.
Furthermore, expanding the observational sample of He/H and O/H ratios in low-metallicity galaxies is crucial. This requires deep spectroscopic observations of a diverse population of dwarf galaxies, including EMPGs, spanning a range of specific star formation rates.  
In addition, the emerging data on high-$z$ galaxies will play a pivotal role in refining our understanding of the He/H-–O/H relationship and its dependence on specific star formation rates.
Together, these efforts will be essential for assessing whether EMPGs are indeed local analogs of the first galaxies and for advancing our understanding of their chemical evolution.


\section{Acknowledgments}
Numerical computations were carried out on the Cray XC50 at the Center for Computational Astrophysics, National Astronomical Observatory of Japan, and 
{\sc SQUID} at the Cybermedia Center, Osaka University, 
as part of the HPCI system Research Project (hp230089, hp240141). 
This work is supported in part by the MEXT/JSPS KAKENHI grant numbers  20H00180, 22K21349, 24H00002, 24H00241 (K.N.), 22KJ0157, 25H00664, 25K01046 (Y.H.), 21J20785 (Y.I.), and 21K03614,21K03633,22H01259,22K03688,24K07095 (T.R.S.). 
This work was supported by JST SPRING, grant number JPMJSP2138 (K.F.).
K.N. acknowledges the travel support from the Kavli IPMU, World Premier Research Center Initiative (WPI), where part of this work was conducted. 
Y.I. is supported by JSPS KAKENHI Grant No. 24KJ0202.
This work was supported by the Global-LAMP Program of the National Research Foundation of Korea (NRF) grant funded by the Ministry of Education (No. RS-2023-00301976).
English language editting was proofread with the assistance of ChatGPT.


%






\appendix

\section{One-zone box model}
\label{sec:appendix_One-zone}
In this section, we present the governing equations and parameters of our one-zone model for the chemical evolution of galaxies, which assumes that the cold ISM is uniformly enriched by metals. Treatment of the chemical enrichment of ISM in this approximation is well established \citep[e.g.][]{Tinsley80, Matteucci_Greggio1986A&A, Matteucci_Francois1989MNRAS, Prantzos1993ApJ, Timmes1995ApJS, Chiappini1997ApJ, Matteucci2001ASSL, Kobayashi2006ApJ, Suzuki_Maeda2018ApJ, Kobayashi20_Origin, Kobayashi_Ferrara2024ApJ}. 
In this study, we mainly follow \citet{Kobayashi_Taylor_2023arXiv}.

The time evolution of the mass fraction $Z_i$ of the $i$th element (H, He, metals) in the gas phase of a one-zone box can be written as follows: 
\begin{equation}
    \label{eq:method_box_model}
    \frac{d\left(Z_i(t){f_\mathrm{gas}}(t)\right)}{dt}=Z_{i,\mathrm{in}}(t)\dot{R}_\mathrm{in}(t)+\dot{E}_{\mathrm{eje},\,i}(t)-Z_i(t)\psi(t)-Z_i(t)\dot{R}_\mathrm{out}(t), 
\end{equation}
where each term on the right-hand side corresponds to the gas inflow fraction rate of the $i$th element, element ejection fraction rate into ISM from SNe, gas mass fraction incorporated into stars during star formation, and the gas outflow fraction rate of the $i$th element from the galaxy by SNe.
Eq.~\ref{eq:method_box_model} is normalized by the total accreted gas mass in a one-zone box model (see Eq.~\ref{eq:method_Minf}).
Here, 
$f_\mathrm{gas}$ is the gas fraction or the total gas mass in the system of a unit mass as a function of time,
$Z_{i,\,\mathrm{in}}$ is the mass ratio of the $i$th element in the accreted gas, 
$\dot{R}_\mathrm{in}$ is the gas accretion fraction rate, 
$\psi$ is the SFR fraction, and
$\dot{R}_\mathrm{out}$ is the gas mass outflow fraction rate.
Note that $f_{\text{gas}}$ is distinct from $\zeta$ in Eq.~\ref{eq:gas_fraction} as it is normalized by the initial gas mass, which does not evolve.
In $\dot{E}_{\mathrm{eje},\,i}$, the total yield is obtained by adding the net yield of each element newly produced by the star to the abundance of each element that the star has at the time of its formation. 
The net yields are based on the values calculated by \citet{Nomoto13, Limongi2018ApJS, Nandal2024AA}.
We explore the varying metallicies of accreted gas with $f_\mathrm{inf}=Z_{i,\mathrm{in}}/Z_i=0, 0.01, 0.1, 1.0$.

The gas mass accretion rate is assumed as
\begin{equation}
    \label{eq:method_Minf}
    \dot{R}_\mathrm{in}(t)=\frac{1}{t_\mathrm{in}}\exp\left(-\frac{t}{t_\mathrm{in}}\right),
\end{equation}
where $\ti$ is the gas accretion timescale. 
The numerator in the first term on the right-hand-side is unity, as it is normalized by the total accreted gas mass. 

The SFR can be written as 
\begin{equation}
    \label{eq:method_SFR}
    \psi(t)=\frac{f_\mathrm{gas}(t)}{\ts},
\end{equation}
where $\ts$ is the star formation timescale.

Although the outflow fraction rate is usually taken as $\dot{R}_\mathrm{out}=\fo \psi$, we adopt SN energy directly, and calculate the outflow fraction rate as follows: 
\begin{equation}
    \label{eq:method_outflow}
    \dot{R}_\mathrm{out}(t)=\dot{e}_\mathrm{SN}(t)\,\frac{100\,\Msun}{10^{51}\,\mathrm{erg}}\times \fo, 
\end{equation}
where the energy injection rate by SN, $\dot{e}_\mathrm{SN}(t)$, is computed as
\begin{equation}
    \label{eq:method_Esn}
    \dot{e}_\mathrm{SN}(t)=\int_0^t\dot{E}_\mathrm{CCSN, SNIa}(t-t_\mathrm{form})\,\psi(t_\mathrm{form})\,dt_\mathrm{form}.
\end{equation}
The term $100\,\Msun/10^{51}\,\mathrm{erg}$ in Eq.~\ref{eq:method_outflow} is based on the assumption that the energy output from a star cluster of $100\,\Msun$ is $10^{51}$\,ergs. 
These values correspond to the outflow mass loading factor, defined as the ratio between the outflow rate and the star formation rate adopted with the SN energy from Table~4 of \citet{Saitoh17}.

Here, $\dot{E}_\mathrm{CCSN, SNIa}(t-t_\mathrm{form})$ is the energy per unit mass emitted per unit time by {\SNII} and SN Ia from individual star clusters, which depends on the SN event rate and the IMF.
Therefore, $\dot{E}_\mathrm{CCSN, SNIa}$ is dependent on the current time $t$ and the star cluster's formation time $t_\mathrm{form}$.
Since $\dot{E}_\mathrm{CCSN, SNIa}(t-t_\mathrm{form})$ is the energy release per unit mass, we multiply by $\psi$ to use the mass at the formation time.

Similarly to Equation \ref{eq:method_Esn}, the element ejection fraction rate of the element $i$ due to stellar evolution can be written as 
\begin{equation}
    \label{eq:method_Meje}
    \dot{E}_{\mathrm{eje},\,i}(t)=\int_0^t \,\psi(t_\mathrm{form}) \,\dot{Y_i}(t-t_\mathrm{form})\,dt_\mathrm{form}, 
\end{equation}
where $\dot{Y_i}(t-t_\mathrm{form})$ represents the mass ejection rate per unit mass of the star cluster per unit time for the $i$th elements, originating from the stellar cluster.
Quantities $\dot{E}_\mathrm{CCSN, SNIa}(t-t_\mathrm{form})$ and $\dot{Y_i}(t-t_\mathrm{form})$ were calculated using {\sc CELib} \citep{Saitoh16, Saitoh17}.

\begin{figure}[ht!]
\begin{center}
    \includegraphics[width=0.95\columnwidth]{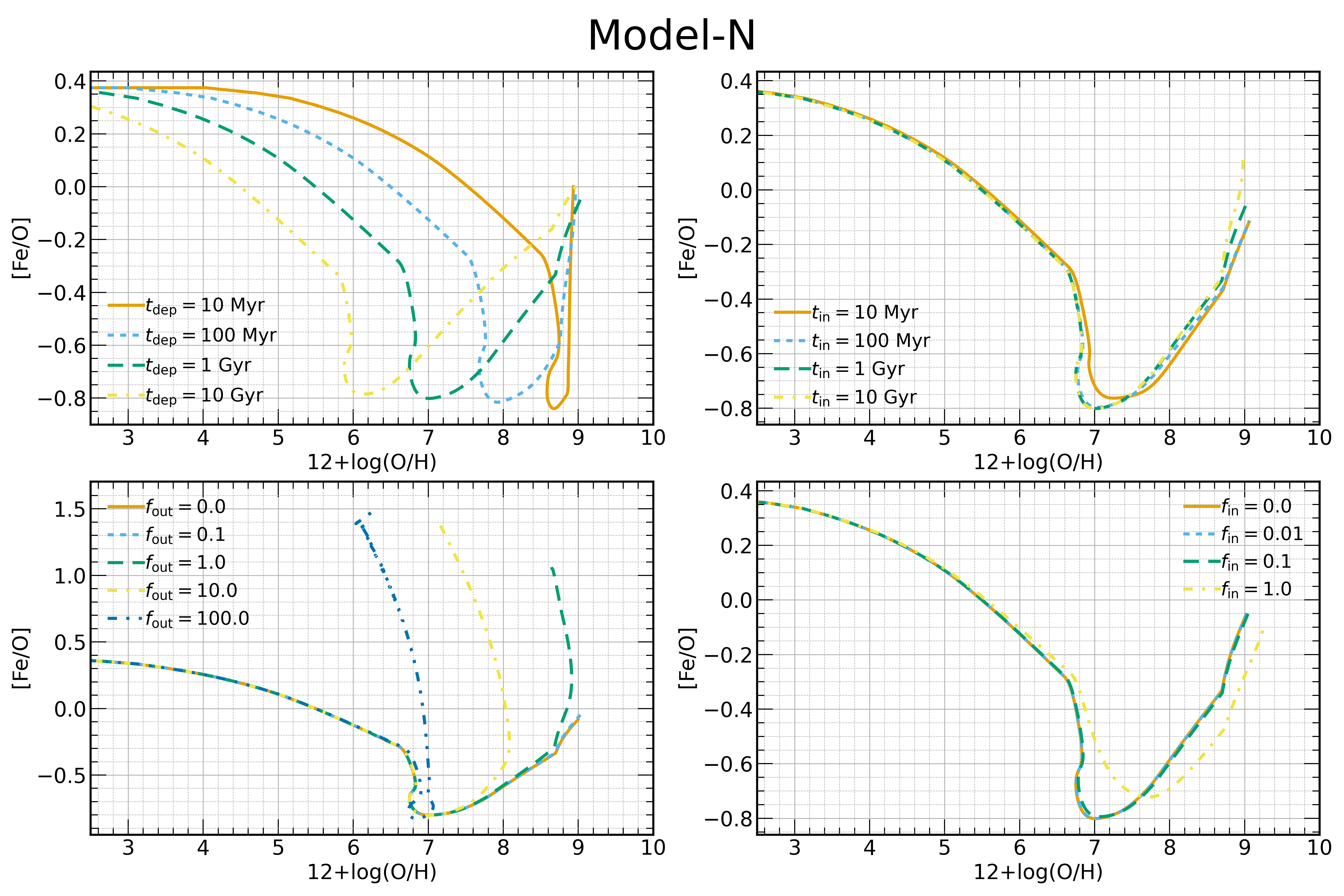}\\
    \includegraphics[width=0.95\columnwidth]{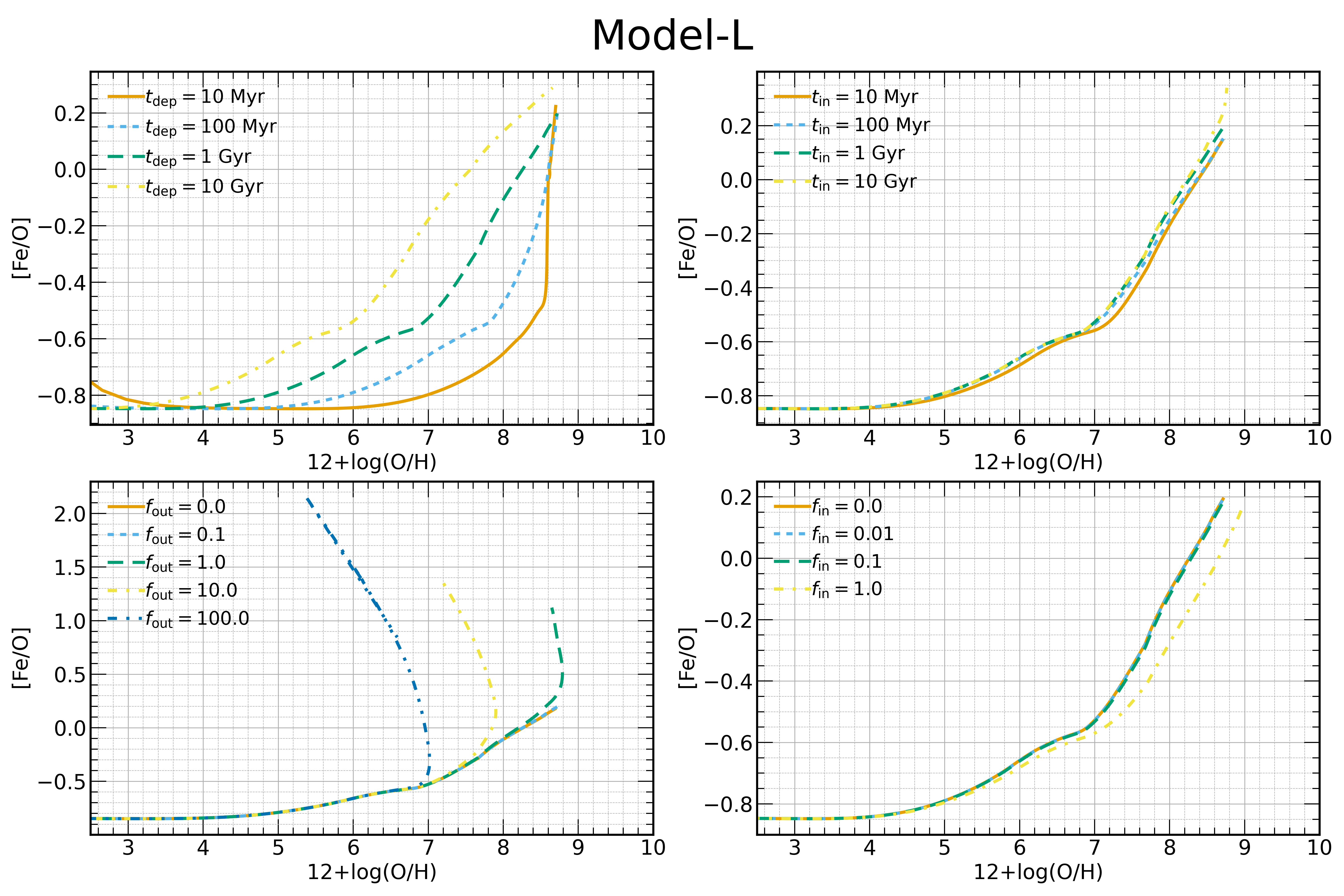}
\end{center}
\caption{Chemical evolution track of Fe/O vs. O/H from our one-zone model. The top 4 panels are for Model-N  \citep{Nomoto13}, and the bottom 4 panels are for Model-L \citep{Limongi2018ApJS}.
Each of the four panels shows the variation in parameters $\ts,\,\ti,\,f_\mathrm{inf}$, and $\fo$  clockwise. 
The following parameters were chosen as fiducial values: $\ts=10^9\,{\rm yrs},\,\ti=10^9\,{\rm yrs},\,f_\mathrm{inf}=0.0$, and $\fo=0.1$. 
\label{fig:Appen_OnezoneModel_FeO_OH}}
\end{figure}

To enhance understanding of the one-zone model calculation, we provide a comparison of fundamental outcomes from different yield models. 
Figure~\ref{fig:Appen_OnezoneModel_FeO_OH} shows the evolutionary track of the one-zone model calculation presented in Figure~\ref{fig:onezone}({\it c,d}). 
The top four panels show Model-N, and the bottom four panels show Model-L. 
Each of the four panels shows the variation in parameters $\ts,\,\ti,\,f_\mathrm{inf}$, and $\fo$  clockwise. 
The pathway of chemical evolution is influenced by the choice of yields, with $\ts$ playing a primary role.
A larger value of $\fo$ leads to an increase in the Fe/O due to SN Ia contributions, accentuating the impact of gas outflow and the most recent metal enrichment.
The evolutionary tracks in the upper four panels begin with a high Fe/O ratio, approximately $0.35$, a consequence of metal enrichment by PISN from {\PopIII} stars (See the red line for $\log \text{Age}\, \mathrm{yr} = 6.4$ in {\it panel (e)} of Figure~\ref{fig:Appen_CElib_OH_HeH_FeO}.) In the \citet{Limongi2018ApJS} yield, on the other hand, metal enrichment by {\SNII} is mainly contributed by the WR star, resulting in low Fe/O values at low metallicities in the one-zone model.

\section{CELib result \lowercase{of} O/H, H\lowercase{e}/H, \lowercase{and} F\lowercase{e}/O}
\label{sec:appendix_CELib}

\begin{figure}[ht!]
\begin{center}
    \includegraphics[width=0.48\columnwidth]{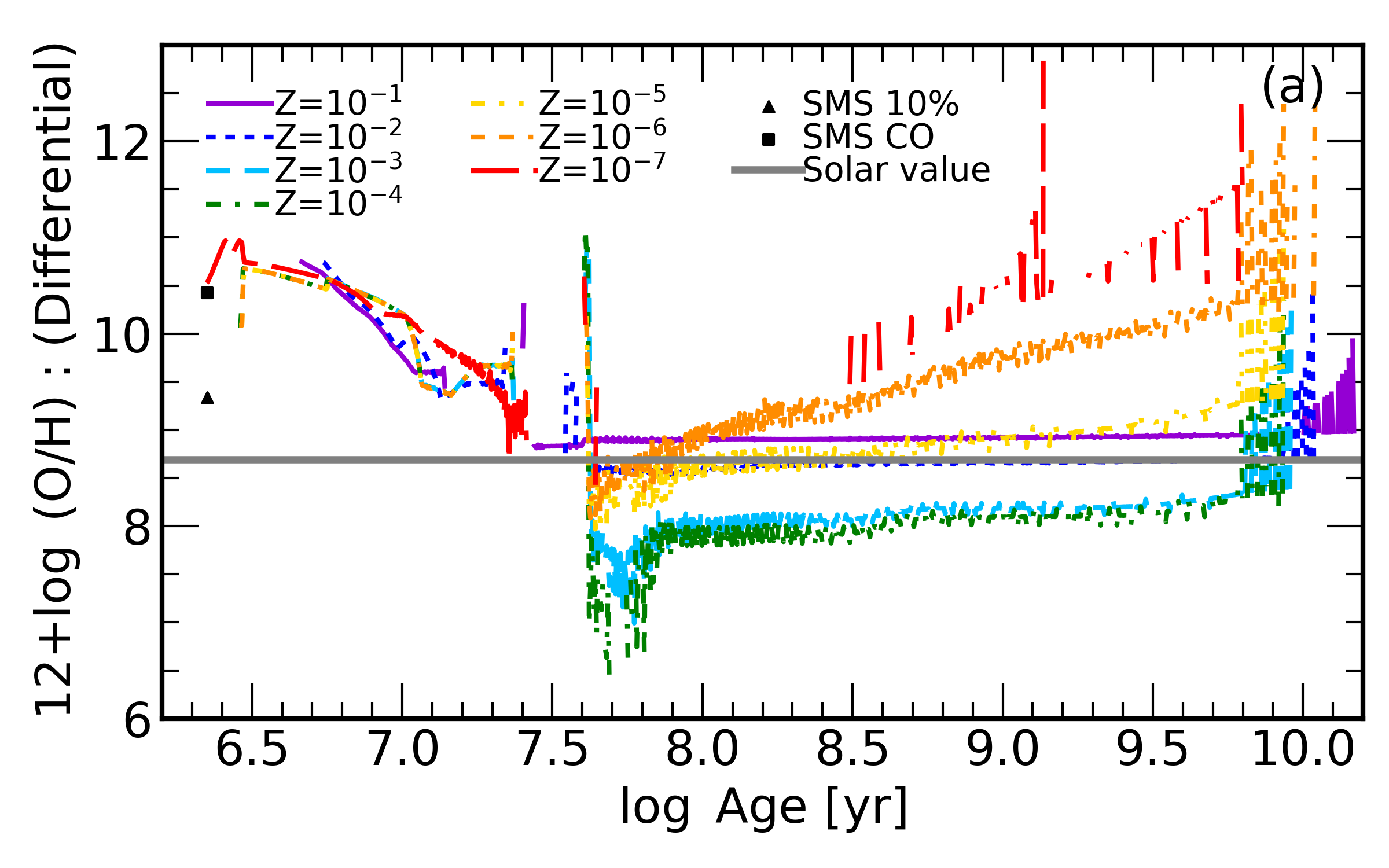}\vspace{1mm}
    \includegraphics[width=0.48\columnwidth]{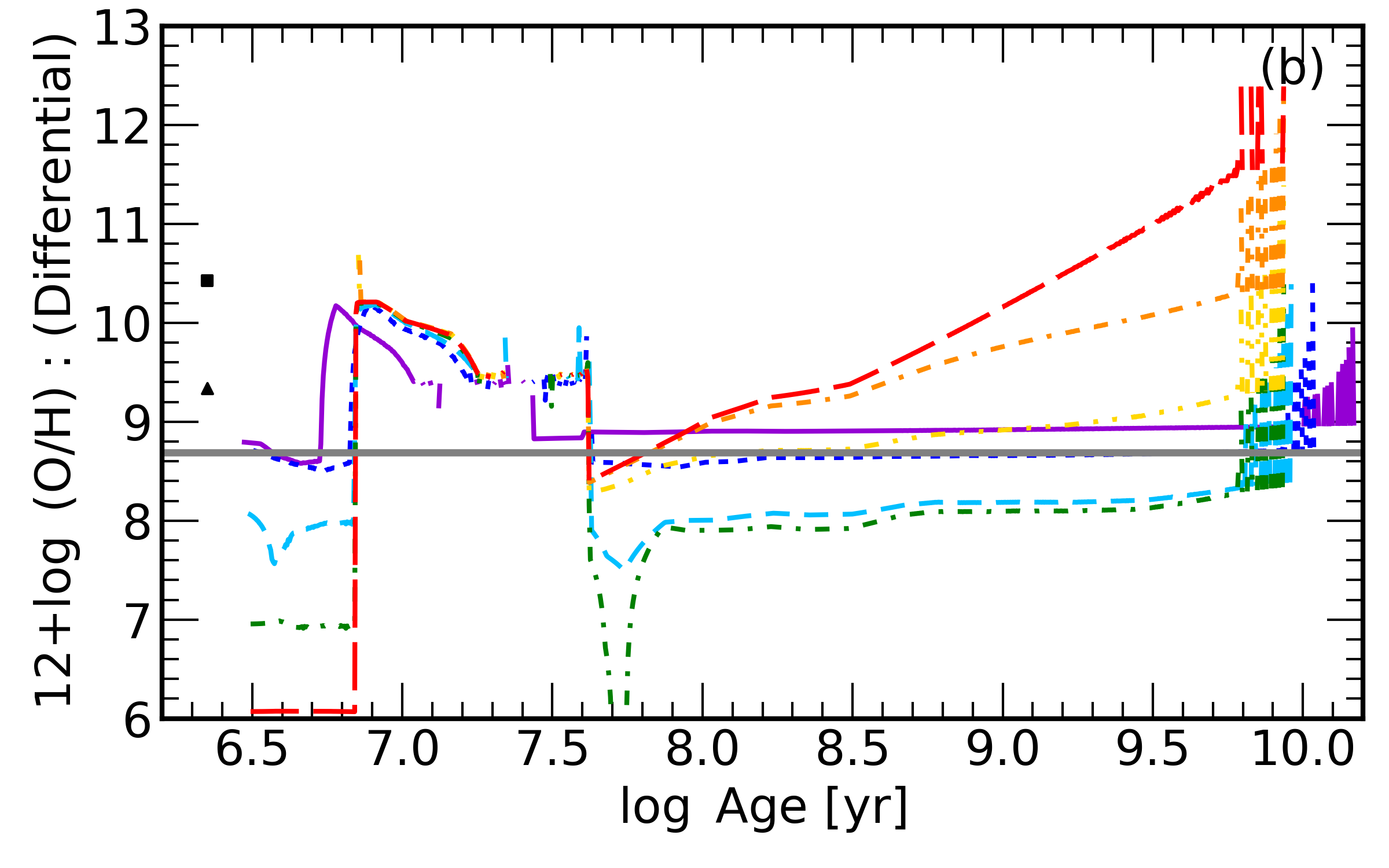}\\
    \includegraphics[width=0.48\columnwidth]{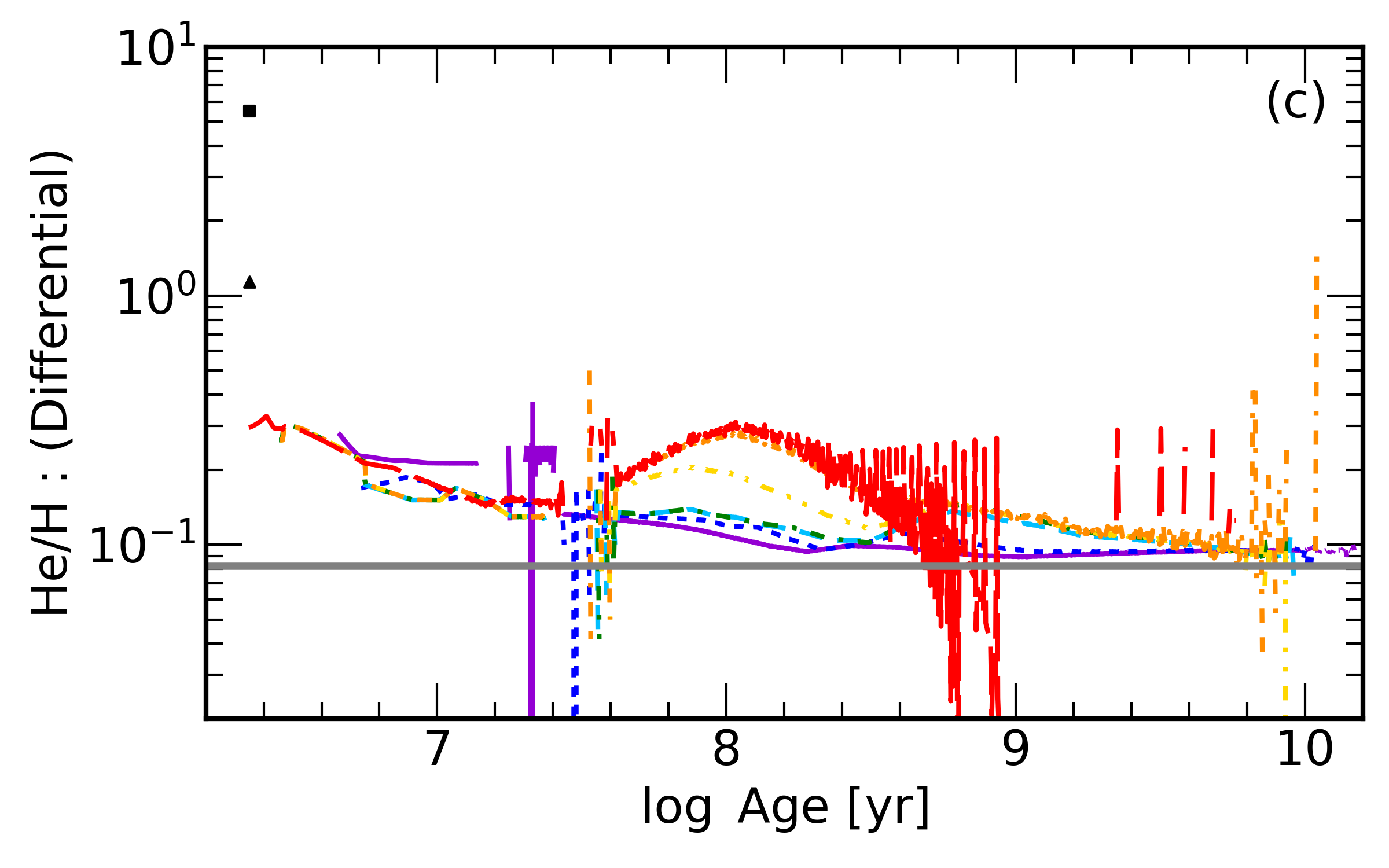}\vspace{1mm}
    \includegraphics[width=0.48\columnwidth]{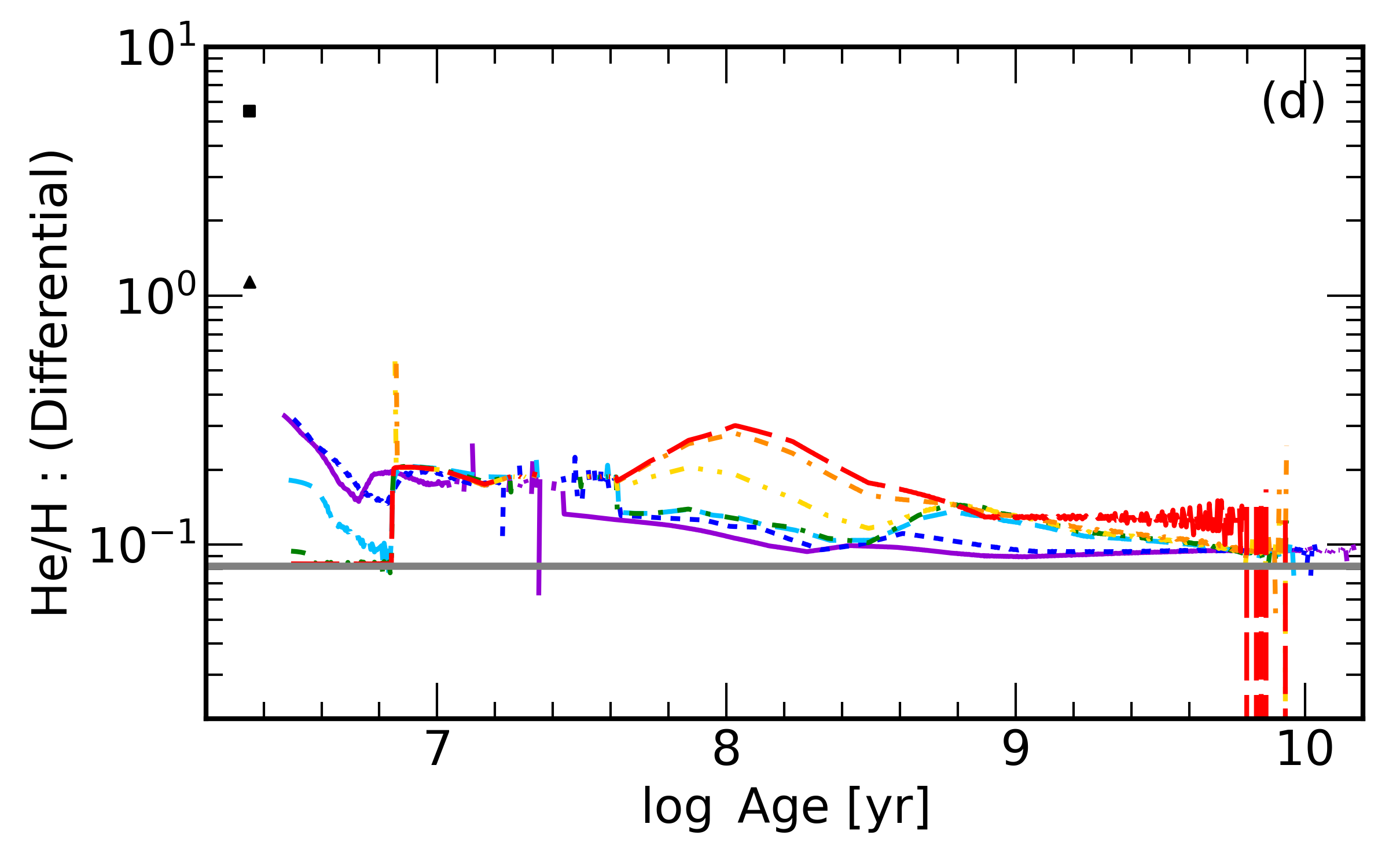}\\
    \includegraphics[width=0.48\columnwidth]{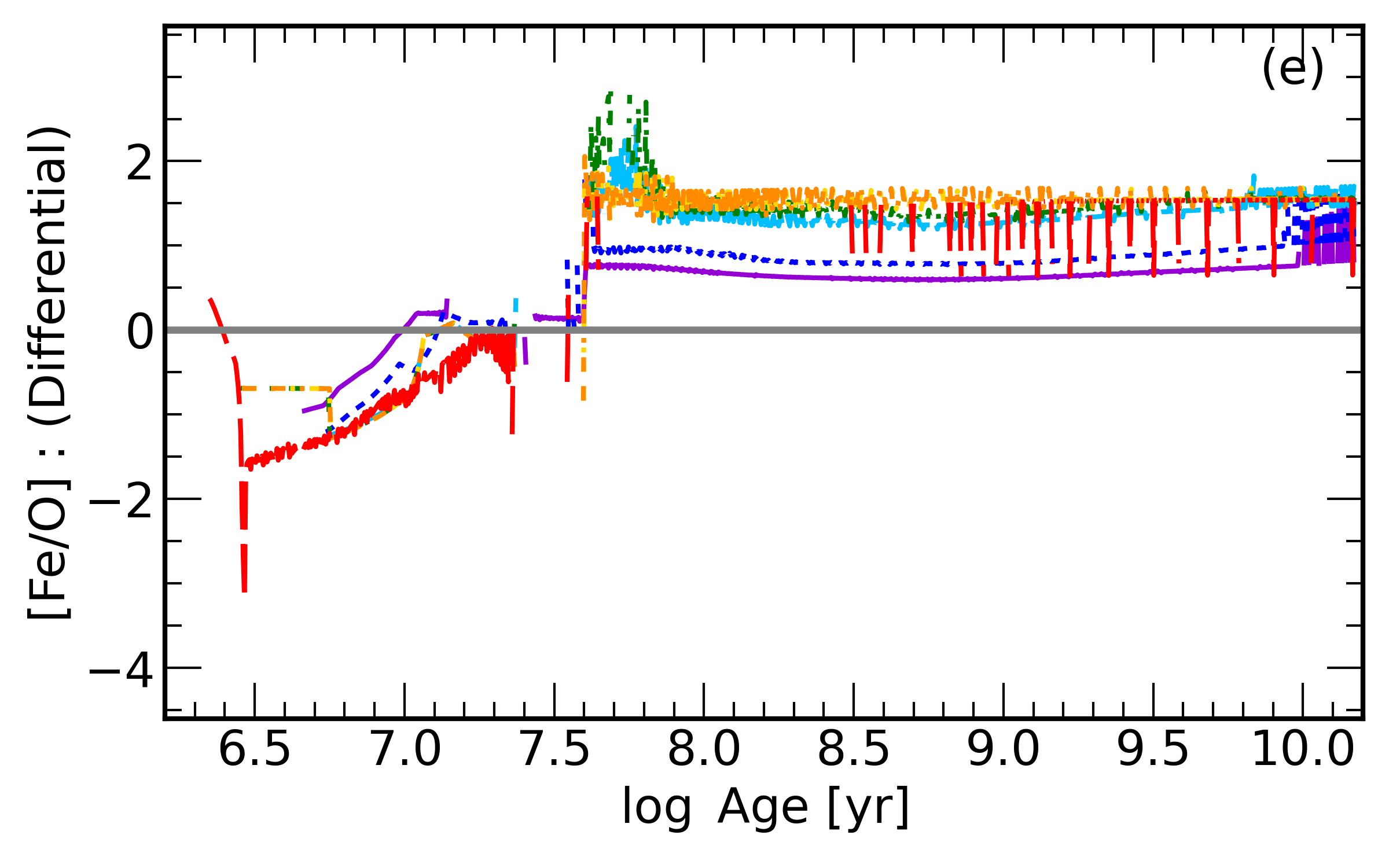}\vspace{1mm}
    \includegraphics[width=0.48\columnwidth]{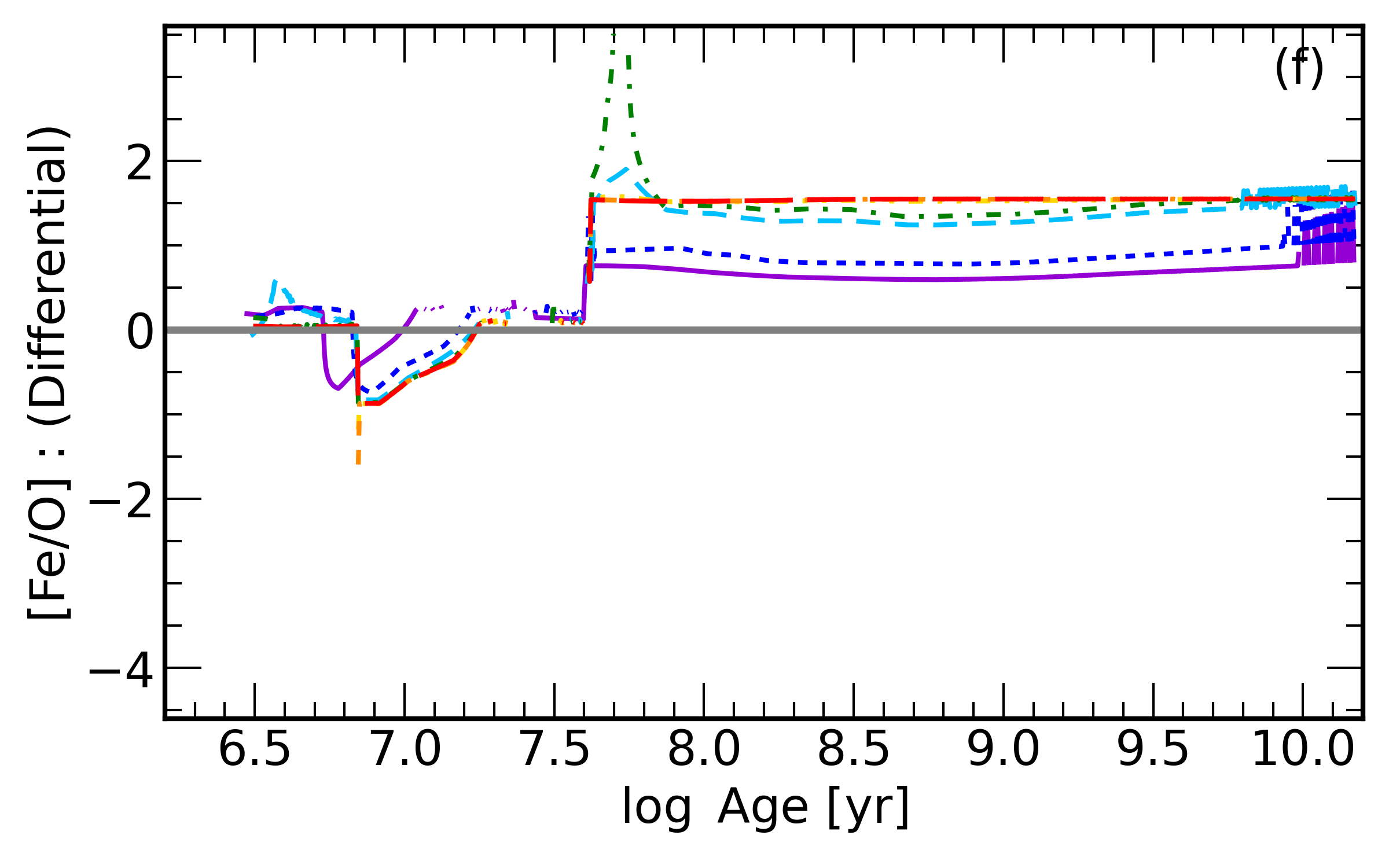}
\end{center}
\caption{Differential output of ejecta's chemical abundance: O/H ({\it panel (a), (b)}), He/H ({\it panel (c), (d)}), and Fe/O ({\it panel (e), (f)}) from {\sc CELib} for an instantaneous burst of a simple stellar population. 
The left column is for Model-N, and the right column is for Model-L. 
The black square shows the yield of SMS for the case where all outer layers above the CO core are ejected for $\Msms=6127\,\Msun$ (Model-N\&N-CO), and the black triangle shows the SMS yield for the case where $10\%$ of the total mass is ejected for $\Msms=6127\,\Msun$ (Model-N\&N-10\%).
The solar abundance is shown as a gray solid line.
\label{fig:Appen_CElib_OH_HeH_FeO}}
\end{figure}

To help understand the results of different yield models, we present the time evolution of O/H ({\it panel (a), (b))}, He/H ({\it panel (c), (d))}, and Fe/O ({\it panel (e), (f)}) emitted from an instantaneous burst of a simple stellar population calculated using {\sc CELib} in Figure~\ref{fig:Appen_CElib_OH_HeH_FeO}. 
In panels ({\it a}), ({\it b}), ({\it c}), and ({\it d}), the SMS yield for the case where $10\%$ of the total mass is ejected for $\Msms=6127\,\Msun$ (Model-N\&N-10\%) is shown by a black triangle, and the SMS yield for the case where all outer layers up to the CO core are ejected for $\Msms=6127\,\Msun$ (Model-N\&N-CO) is shown by a black square. 
At $t=10^{7.6}\, \mathrm{yr}$ in panels ({\it e}) and ({\it f}), Fe release by SNIa begins to occur, increasing Fe/O in both panels.
The high Fe/O ([Fe/O]$>0$) at $\sim 10^{6.5}\, \mathrm{yr}$ for $Z=10^{-7}$ ({\PopIII}) case in the panel ({\it e}) is due to metal enrichment by PISN.  Additionally, in panel ({\it f}), a high [Fe/O] ([Fe/O]$>0$) does not appear because the massive star collapses directly to BH.

\section{Gas fraction}
\label{sec:appendix_gasfrac}

\begin{figure}[ht!]
\begin{center}
    \includegraphics[width=0.48\columnwidth]
    {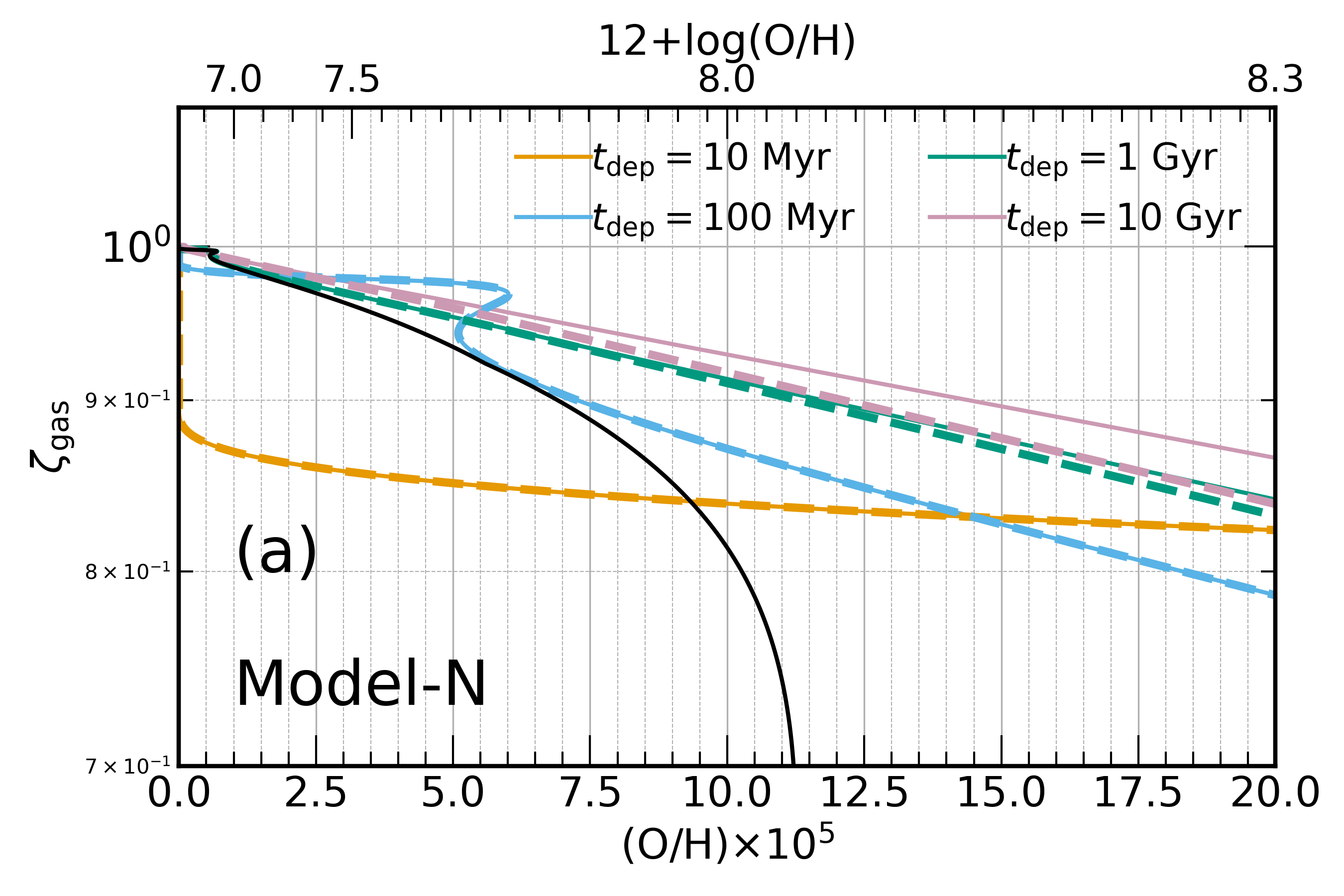}
    \includegraphics[width=0.48\columnwidth]
    {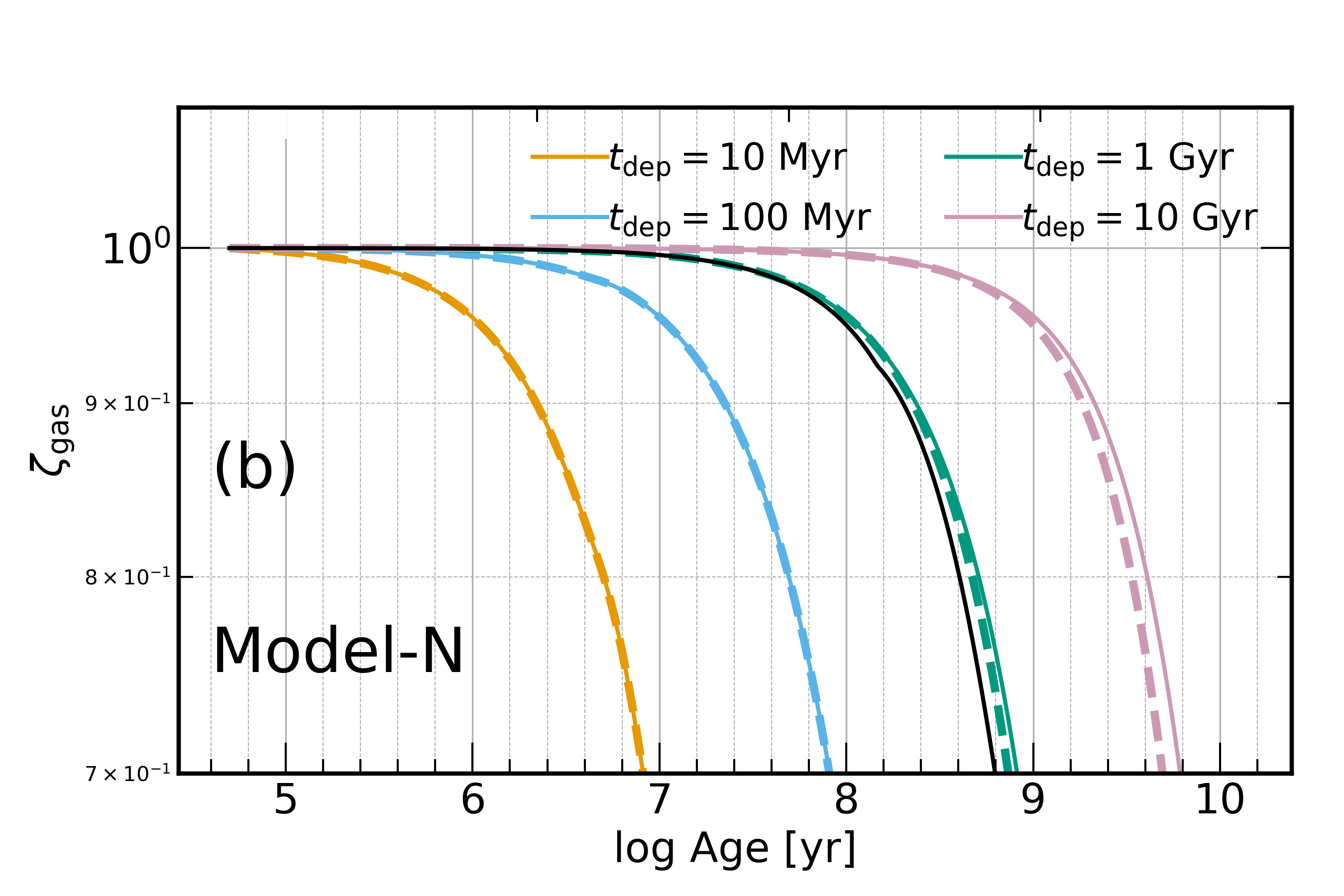}
\end{center}
\caption{
{\it Panel (a)}: $(\mathrm{O/H})\times10^5$--$\zeta_\mathrm{gas}$ relationship obtained from the one-zone model calculations of Model-N. 
The orange, blue, green, and red lines represent $\ts=10$ Myr, 100 Myr, 1 Gyr, and 10 Gyr, respectively (same as Figure~\ref{fig:Nom_Lim_onezone}). The solid lines indicate the results using CCSN, SNIa, and AGB yields, while the dashed lines indicate the results using only CCSN yields. The other parameters are set to $\ti=1$ Gyr, $\fo=0.1$, and $f_{\mathrm{inf}}=0.1$. The black line represents the evolutionary track for the case of $(\ts[\mathrm{yr}], \ti[\mathrm{yr}], \fo, f_\mathrm{inf}) = (10^9, 10^{10}, 10, 0.01)$ same as Figure~\ref{fig:onezone}.
{\it Panel (b)}: $\log \mathrm{Age}$ [yr] --$\zeta_\mathrm{gas}$ relationship obtained from the one-zone model calculations of Model-N. }
\label{fig:appen_gasfrac}
\end{figure}

Figure~\ref{fig:appen_gasfrac} shows the correlation between $\zeta_\mathrm{gas}$ and the metallicity ({\it panel a}) and its time evolution ({\it panel b}). 
The plotted models are the same as those in Figure~\ref{fig:Nom_Lim_onezone}.
The black line represents the evolutionary track for the case of $(\ts[\mathrm{yr}], \ti[\mathrm{yr}], \fo, f_\mathrm{inf}) = (10^9, 10^{10}, 10, 0.01)$ same as the black line in Figure~\ref{fig:onezone}.
In panel {\it (a)}, the blue curve ($\ts = 100$ Myr) shows a decrease in O/H around $(\text{O/H}) \times 10^5 = 5$. This is due to the transition from {\PopIII} to {\PopII} stars.
The Model-N assumes that {\PopIII} stars release more oxygen than {\PopII} stars. As the dominant contributors of oxygen transition from {\PopIII} to {\PopII}, the oxygen ejection decreases. 
Once the stellar mass of {\PopII} stars becomes sufficiently large to produce and release oxygen efficiently, the metallicity starts to increase again.
A similar trend can be seen for other curves, such as the black curve ($\ts = 1$ Gyr), where the transition occurs around $(\text{O/H}) \times 10^5 = 0.6$ for the same reason.

\section{Outflow of our one-zone model}
\label{sec:appendix_outflow}

\begin{figure}[ht!]
\begin{center}
    \includegraphics[width=0.48\columnwidth]{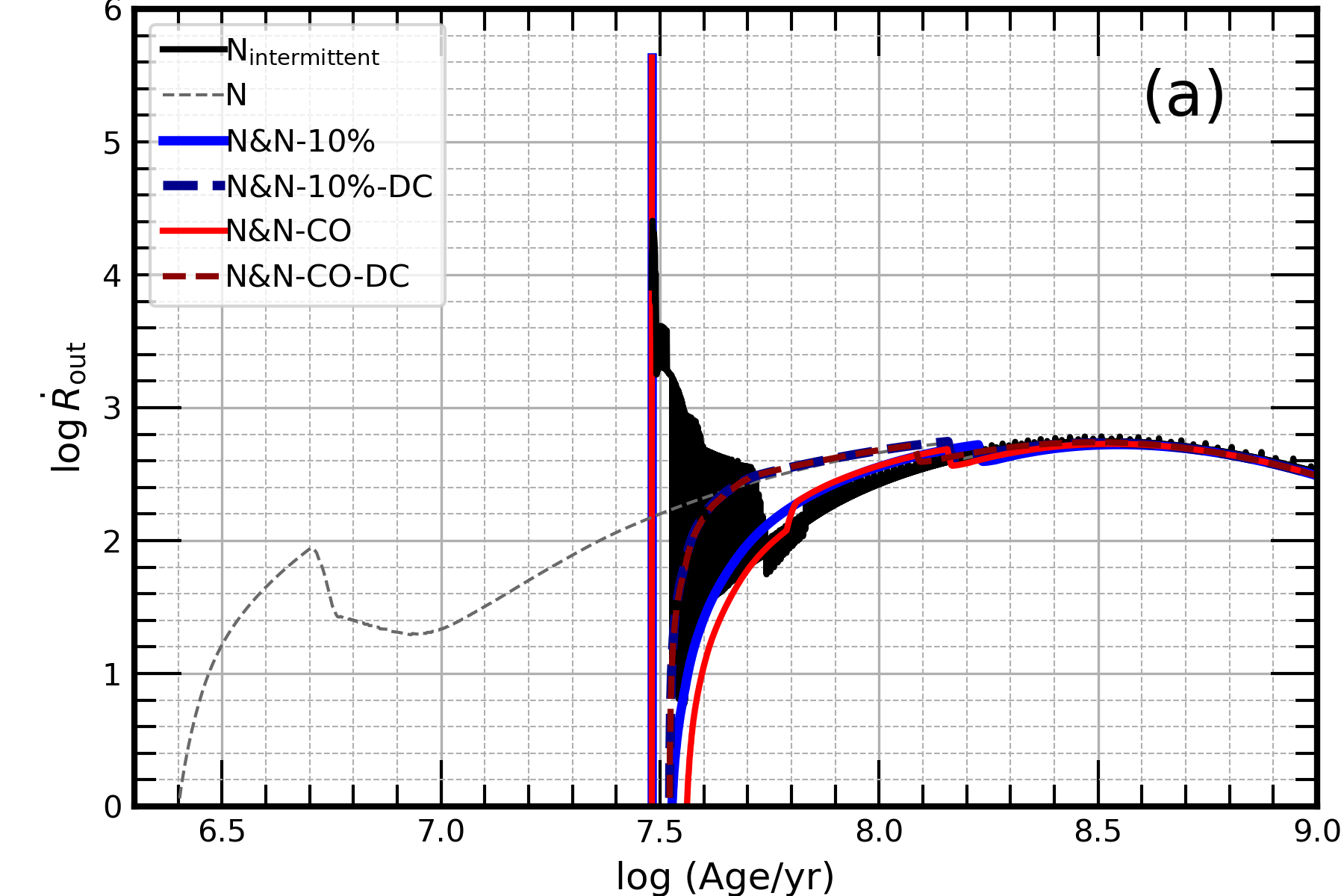}
    \includegraphics[width=0.48\columnwidth]{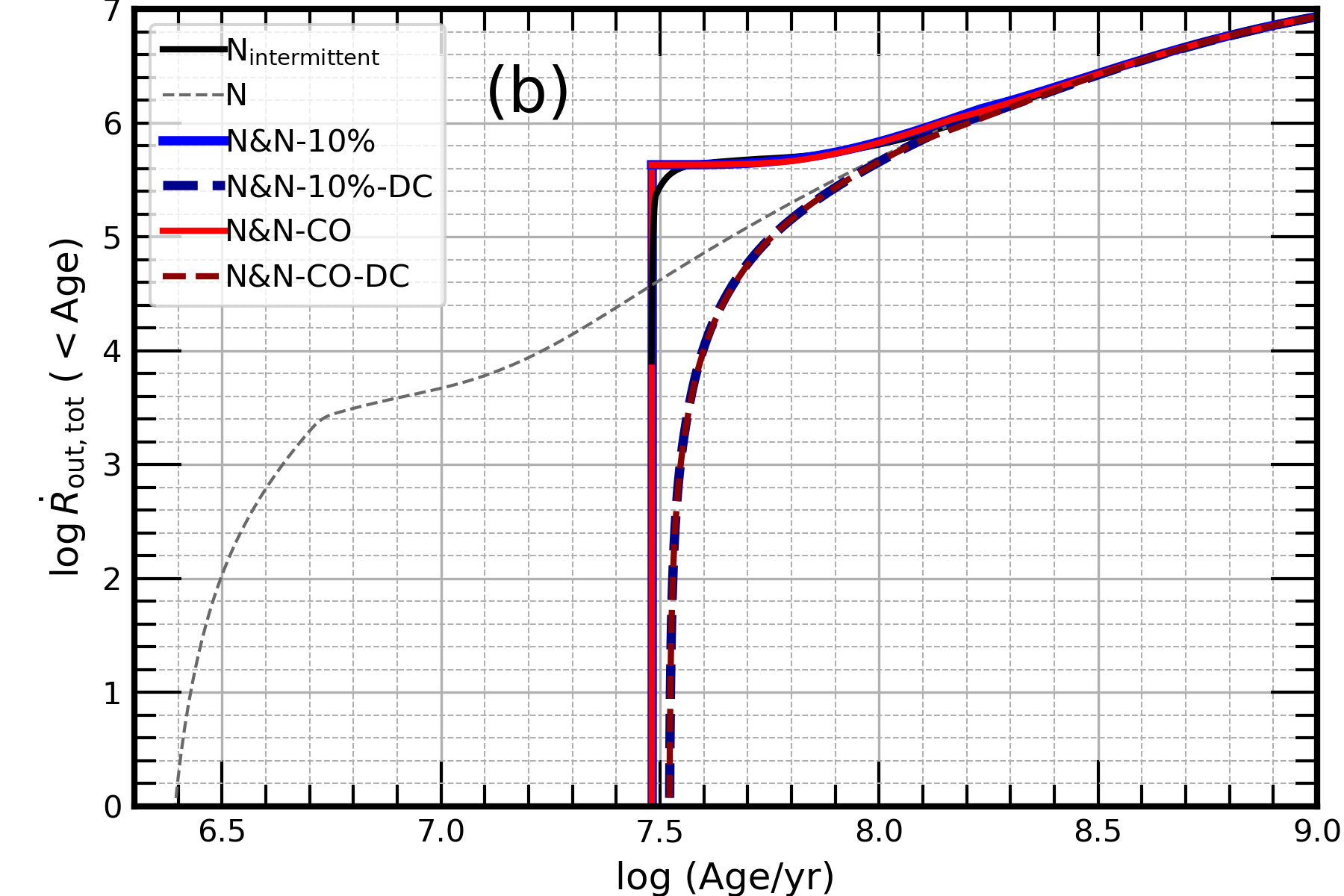}
    
\end{center}
\caption{Time evolution of gas outflow fraction rate $\dot{R}_\mathrm{out}$ for the Model-N, $\mathrm{N}_\mathrm{int}$, $\mathrm{N}_\mathrm{int}$, and N\&N series.
Panel ({\it a}) shows the values at each timestep, while panel ({\it b}) displays the cumulative values.
Here, $\ts=10^9 \,\mathrm{yr}$, $\ti=10^9 \,\mathrm{yr}$, $\fo=10$, and $f_\mathrm{inf}=0$ are used.
}
\label{fig:Appen_time_Mout}
\end{figure}

Figure~\ref{fig:Appen_time_Mout} ({\it a}) shows the time evolution of gas outflow fraction rate $\dot{R}_\mathrm{out}$ for the Model-N, $\mathrm{N}_\mathrm{int}$, and N\&N series. 
The line styles follow those in Figure~\ref{fig:time_evolution_onezoneSMS}.
The panel ({\it b}) shows the time evolution of the cumulative $\dot{R}$. 
Model-$\mathrm{N}_\mathrm{int}$, which forms stars discretely, differs from Model-N in terms of {\PopIII} star mass. In Model-$\mathrm{N}_\mathrm{int}$, star formation is delayed until a star with $\Msms$ is formed to match the SMS case, whereas in Model-N, {\PopIII} stars are formed according to Eq.~\ref{eq:method_SFR}. The resulting metal enrichment triggers the formation of {\PopII} stars.
For the {\PopIII} case, the Susa IMF is assumed, leading to the formation of a larger number of massive stars. 
Consequently, $\dot{E}_\mathrm{CCSN}$ in Eq.~\ref{eq:method_Esn} is higher for {\PopIII} than for {\PopII}. 
As a result, the amount of outflow per unit stellar mass is also stronger in Model-$\mathrm{N}_\mathrm{int}$ compared to Model-N, as shown in the right panel of Figure~\ref{fig:Appen_time_Mout}.

\bibliography{Fukushima}{}
\bibliographystyle{aasjournal}



\end{document}